\documentclass[12pt]{article} 
\usepackage[margin=0.8in]{geometry}
\usepackage{graphicx,caption}
\usepackage{natbib}
\usepackage{ASSmarco}
\usepackage{amsmath}
\usepackage{wasysym}
\usepackage{authblk}
\usepackage{hyperref}
\usepackage{multicol}
\usepackage{gensymb}
\let\origappendix\appendix 
\renewcommand\appendix{\clearpage\pagenumbering{roman}\origappendix}
\usepackage[usenames, dvipsnames]{color}
\usepackage{multirow}
\usepackage [autostyle, english = american]{csquotes}
\MakeOuterQuote{"}

\usepackage{pdflscape}
\usepackage{listings}
\usepackage{color}
\usepackage{standalone}

\newenvironment{Figure}
  {\par\medskip\noindent\minipage{\linewidth}}
  {\endminipage\par\medskip}
  
\newenvironment{Table}
  {\par\medskip\noindent\minipage{\linewidth}}
  {\endminipage\par\medskip}

\graphicspath{{/home/sebastiano/Documents/Data/paper_A_dection_of_SgrA_/submitted_images/}}

\begin{document}

\title{A Detection of Sgr~A* in the far infrared} 
\author[1]{Sebastiano D. von Fellenberg}
\author[1]{Stefan Gillessen}
\author[1]{Javier Graci\'{a}-Carpio}
\author[2,3]{Tobias K. Fritz}
\author[1]{Jason Dexter}
\author[1]{Michi Baub\"ock}
\author[1]{Gabriele Ponti}
\author[1]{Feng Gao}
\author[1]{Maryam Habibi}
\author[1]{Philipp M. Plewa}
\author[1]{Oliver Pfuhl}
\author[1]{Alejandra Jimenez-Rosales}
\author[1]{Idel Waisberg}
\author[1]{Felix Widmann}
\author[1]{Thomas Ott}
\author[1]{Frank Eisenhauer}
\author[1,5]{Reinhard Genzel}
\affil[1]{Max Planck Institute for Extraterrestrial Physics, 85748, Garching, Germany}
\affil[2]{Instituto de Astrofisica de Canarias, calle Via Lactea s/n, E-38205 La Laguna, Tenerife, Spain}
\affil[3]{Universidad de La Laguna, Dpto. Astrofisica, E-38206 La Laguna, Tenerife, Spain}
\affil[5]{Astronomy \& Physics Departments, University of California, Berkeley, CA 94720, USA}

\maketitle


\begin{abstract}
\noindent We report the first detection of the Galactic Centre massive black hole, Sgr~A*, in  the far infrared. Our measurements were obtained with PACS on board the \emph{Herschel} satellite at $100~\mathrm{\mu m}$ and $160~\mathrm{\mu m}$.
While the warm dust in the Galactic Centre is too bright to allow for a direct detection of Sgr~A*, we measure a significant and simultaneous variation of its flux of $\Delta F_{\nu\widehat{=}160 ~\mathrm{\mu m}} = (0.27\pm0.06)~\mathrm{Jy}$ and $\Delta F_{\nu\widehat{=}100 ~\mathrm{\mu m}}= (0.16\pm0.10)~\mathrm{Jy}$ during one observation. The significance level of the $160 ~\mathrm{\mu m}$ band variability is $4.5\sigma$ and the corresponding $100 ~\mathrm{\mu m}$ band variability is significant at $1.6\sigma$. We find no example of an equally significant false positive detection. Conservatively assuming a variability of $25\%$ in the FIR, we can provide upper limits to the flux. Comparing the latter with theoretical models we find that 1D RIAF models have difficulties explaining the observed faintness. However, the upper limits are consistent with modern ALMA and VLA observations. Our upper limits provide further evidence for a spectral peak at $\sim 10^{12} ~ \mathrm{Hz}$ and constrain the number density of $\gamma \sim 100$ electrons in the accretion disk and or outflow.
\end{abstract}

\begin{multicols}{2}
\section{Introduction}
The Galactic Centre massive black hole, Sgr~A*, and its accretion flow have long been established as a one of kind laboratory that grants access to exceptional physical phenomena \citep{2010RvMP...82.3121G}. 
The emission stemming from the accretion flow (and or outflow) has been measured throughout many parts of the electromagnetic spectrum ranging from the radio \citep{2001ARAA..39..309M}, the mm  \citep{2003ApJ...586L..29Z}), the sub-mm \citep{1998ASPC..144..323F} and the NIR \citep{2003Natur.425..934G} to the X-ray \citep{2001AAS...199.8503B} regime.

These measurements make up the spectral energy distribution (SED) of Sgr~A*. The power, variability and spectral slope vary substantially throughout the SED. Reflecting that, the different parts of the SED have been given different phenomenological names: the radio part is `flat' (i.e. the flux is approximately log-constant, \citealt{1997ApJ...490L..77S}) and thus dubbed the flat radio tail; the spectral slope increases and peaks in the mm to sub-mm domain of the SED \citep{1998ASPC..144..323F}. This peak has sometimes been referred to as the `sub-mm bump'.

At wavelengths shorter than $1~\mathrm{mm}$ the observation of Sgr~A* becomes more difficult due to obscuration from the atmosphere. Sgr~A* has been observed with the Caltech Submillimeter Observatory (CSO) at wavelengths down to $350~\mathrm{\mu m}$ (e.g. \cite{2009ApJ...706..348Y}). \cite{2016ApJ...825...32S} report `highly significant variations' of the deviation from the mean flux and a `minimum time-averaged flux density' of $\langle\Delta F_{\nu \widehat{=}250\mathrm{\mu m}}\rangle=0.5~\mathrm{Jy}$ using the SPIRE instrument on-board Herschel. 

At even shorter wavelengths only upper limits exist, until Sgr~A* reappears in the NIR, where its variable outbursts are frequently recorded \citep{2003Natur.425..934G,2009ApJ...698..676D}. 
In the optical and UV regime, dust extinction makes observations of Sgr~A* impossible. 

In the X-ray regime, both a variable as well as a constant flux component are observed. The constant X-ray flux has a spatial extension consistent with the Bondi radius ($\sim 1" ~ \widehat{=} ~ 10^5$ Schwarzschild radii) of Sgr~A* \citep{2003ApJ...591..891B,2006ApJ...640..319X}. The variable flux is thought to originate from the innermost part ($\approx 10~\mathrm{R_S}$) of the accretion flow \citep{2014ApJ...786...46B}.

Sgr~A* is a variable source at all observable wavelengths \citep{2010RvMP...82.3121G}. However, it is not clear whether the variability in different spectral regimes is physically connected \citep{2014MNRAS.442.2797D}. It has been established that all X-ray flares are accompanied by a NIR flare. But the converse is not true \citep{2009ApJ...698..676D}.

Both the NIR \citep{2009ApJ...691.1021D} and the (sub)-mm variability shows red noise characteristics.  The sub-mm emission has a  characteristic time scale of $\tau=8\mathrm{h}$ \citep{2014MNRAS.442.2797D}. 

The fractional variability increases throughout these wavelength regimes. In the cm, mm and sub-mm regime the variability is in the order of a few tens of percent. In the NIR regime the range of the variability increases and is of the order of a few hundred percent. In the X-ray regime it is yet a magnitude larger \citep{2010RvMP...82.3121G}.

Based on these observational constraints, the emitting material has been modeled by two broad classes of models: Radiatively Inefficient Accretion Flow (RIAF) models and jet models. Both types of model can explain the observed SED.

In RIAF models, two populations of electrons exist: a thermal population producing the emission in the sub-mm and mm regime and an accelerated fraction of (non-thermal) power law electrons producing the flat radio tail at longer wavelengths \citep{2003ApJ...598..301Y,2004ApJ...606..894Y}. 

In such an accretion flow the released energy is advected inwards rather than radiated away (and thus the flow is radiatively inefficient). The accretion flow has a geometrically thick and optically thin disk \citep{1977ApJ...214..840I,1982Natur.295...17R,1994ApJ...428L..13N,2003ANS...324..453Y,2014ARA&A..52..529Y}. 

In the jet models, relativistic, optically thick and symmetric jets are responsible for the radio and mm emission as well as the constant X-ray flux. The jet model is motivated phenomenologically from the observed jets in many known low-luminosity active galactic nuclei such as M81 or NGC4258 \citep{2000A&A...362..113F}.

In this context, the emission is produced either by the bulk accretion flow (e.g.,  \citealt{2009ApJ...706..497M}, \citealt{2010ApJ...717.1092D}, \citealt{2012ApJ...755..133S}) or at the jet wall (\citealt{2013A&A...559L...3M}, \citealt{2015ApJ...812..103C}). The latter scenario naturally results from the expected preferential heating of electrons in magnetized regions (\citealt{2010MNRAS.409L.104H}, \citealt{2016arXiv161109365R}) and reproduces the radio spectrum with purely thermal electrons. In the former scenario, an additional non-thermal component is required (\citealt{2000ApJ...541..234O}, \citealt{2003ApJ...598..301Y}, \citealt{2006ApJ...638L..21B}, \citealt{2017MNRAS.466.4307M}, \citealt{2017MNRAS.470.2367C}). Several of these works are also time-dependent and can produce the observed mm and to some extent the NIR variability (\citealt{2009ApJ...703L.142D}, \citealt{2012ApJ...746L..10D}, \citealt{2013MNRAS.432.2252D}, \citealt{2015ApJ...812..103C}, \citealt{2016arXiv161109365R}). However, no simulation so far produces large X-ray flares (but see \citealt{2016ApJ...826...77B}, which can reproduce the X-ray/NIR observations by the stochastic injecting non-thermal electrons).

Until now, due to the obscuration by the atmosphere, as well as the technical challenges far-infrared (FIR) detectors pose, the FIR regime of Sgr~A* has not been constrained. This regime is important though, as the luminosity of the accretion flow is thought to turn over in this regime. Being able to constrain the SED in the FIR would make it possible to narrow down the many degeneracies still present in theoretical models of the accretion flow. This is especially interesting in the context of 3D simulations, where the number of free parameters allow a wide range of simulations to fit the data.

In this paper, we present novel Herschel\footnote{\emph{Herschel} is an ESA space observatory with science instruments provided by European-led Principal Investigator consortia and with important participation from NASA.} FIR measurements and a first detection of Sgr~A* at $\lambda = 100 ~\mathrm{\mu m}$ and $\lambda = 160~ \mathrm{\mu m}$. In section \ref{OandR} we present the observations and the data reduction. In section \ref{R} we describe the results. These are discussed in section \ref{discussion}. Finally, we summarize our results in section \ref{SaO} and give an outlook.


\section{Observations and Reductions}
\label{OandR}
Our observations consist of five slots of coordinated observations in March 2012 with the PACS instrument \citep{2008SPIE.7010E..05P} onboard the ESA Herschel Space Observatory \citep{2010A&A...518L...1P}, parallel X-ray observations with the Chandra\footnote{Obsids: 13856,13857 \& 14413} \citep{2000SPIE.4012....2W} and XMM-Newton\footnote{Obsids: 0674600601, 0674600701, 0674601101, 0674600801 \& 0674601001} \citep{2001A&A...365L...1J,2001A&A...365L..18S} observatories, and the near-infrared NACO camera \citep{2003SPIE.4841..944L,2003SPIE.4839..140R} mounted on UT4 at the VLT observatory. The observing times and the exposure times for the individual instruments are listed in Table \ref{ObsTime}. 

The PACS camera had two bolometer arrays: one operating at either $70~\mathrm{\mu m}$ or $100~\mathrm{\mu m}$ (the `blue' and `green' bands respectively) and one operating at $160~\mathrm{\mu m}$ (the `red' band). Three of the five slots used the blue band filter and two the green band filter (March 17 and March 19).  The parallel X-ray ($2-10~\mathrm{keV}$) and NIR ($K$ and $L'$ band) observations were scheduled to observe as much in parallel as possible.

We used the scan observing mode for PACS. We chose a scanning pattern  that creates images with a total exposure of 10 minutes each. 
The X-ray observations are binned to $300$ seconds exposures; the NACO $K$ and $L'$ band observations have a cadence of 4 and 1 minute respectively. When feasible the NIR filters of NACO were switched to allow for quasi-parallel $K$ and $L'$ observations.  

\end{multicols}
\begin{Table}
\centering
\resizebox{\columnwidth}{!}{
\begin{tabular}{|c||c|c|c|c|c||c|}
\hline 
Instrument & 03/13/2012 & 03/15/2012 & 03/17/2012 & 03/19/2012 & 03/21/2012 &Exposure time / Bins\\ 
\hline 
\hline
PACS&05:13 - 13:05&05:03 - 12:55&05:17 - 13:09&05:08 - 13:00&05:06 - 12:58&10 min\\ 
\hline 
NACO - K& - &08:04 - 8:49&08:18 - 09:47&7:35 - 10:05&07:19 - 10:07&4 min\\ 
\hline 
NACO - L& - &09:36 - 10:15&5:48 - 10:04&08:04 - 10:07&06:08 - 10:08&1 min\\ 
\hline 
XMM-Newton&03:52 - 09:23&04:47 - 08:42&02:30 - 09:50&03:52 - 09:48&03:31 - 09:41&5 min\\ 
\hline 
Chandra&-&08:45 - 19:45 &08:58 - 19:49&-&06:46 - 11:12& 5 min\\ 
\hline 
\end{tabular} 
}
\captionof{table}{Observation time in UT for all available instruments.}
\label{ObsTime}
\end{Table}
\begin{multicols}{2}

A quick look at the images obtained with the standard pipeline reveals that Sgr~A* is not readily seen. There is, however, bright thermal dust emission from the circumnuclear disk (CND, \citealt{2011AJ....142..134E}, see Figure \ref{rgbImage} and Appendix \ref{medianMaps}). Subtracting this constant emission from the individual exposures allows us to look for a variable component of Sgr~A*. The subtraction creates a data cube of $40$ residual 
maps per observation.

The residual maps are dominated by systematic artefacts which make it, at first glance, impossible to detect Sgr~A* as a variable source.

To remove these systematic errors we chose an approach in which we remove the respective dominant artefact step-by-step.

In the following we describe how we obtained the images (subsec. \ref{standardReduction}) and the residual maps (subsec. \ref{advancedReduction}).

\subsection{Standard reduction}
\label{standardReduction}
To create the images, we use the HIPE pipeline \citep{2010ASPC..434..139O} and the JScanam map maker \citep{2015arXiv151203252G}. We keep the standard settings and only change the source masking parameter. This ensures that, in source regions, JScanam's algorithm removes the 1/f noise based on averages. This protects the real signal of a source from being removed. 
We tune this parameter such that the source masks do not cover too much area (a good value for the coverage being $\sim30\%$, J. Graci\'{a}-Carpio priv. comm.). Additionally, we create a square source mask which covers Sgr~A* over an area of $6"$, $7"$ and $12"$ ($4\times4~ \mathrm{px}$) in accordance with Herschel's beam sizes at the three wavelengths. This creates $40$ individual images for each observation.

\subsection{Improved reduction of maps}
\label{advancedReduction}
Here, we detail the steps beyond the standard reduction which enable us to reach a sensitivity of $\sim0.1\mathrm{Jy/beam}$.

\subsubsection{Pointing correction}
\label{pointing}
The Herschel satellite experienced absolute pointing offset errors on the order of $1"$ to $2"$ \citep{2014ExA....37..453S}. This creates strong, spatially correlated patches in the residual maps at regions of high intensity. 

\begin{Figure}
\centering
	\includegraphics[width=\textwidth,keepaspectratio,trim={0.cm 0cm 0cm 0cm},clip]{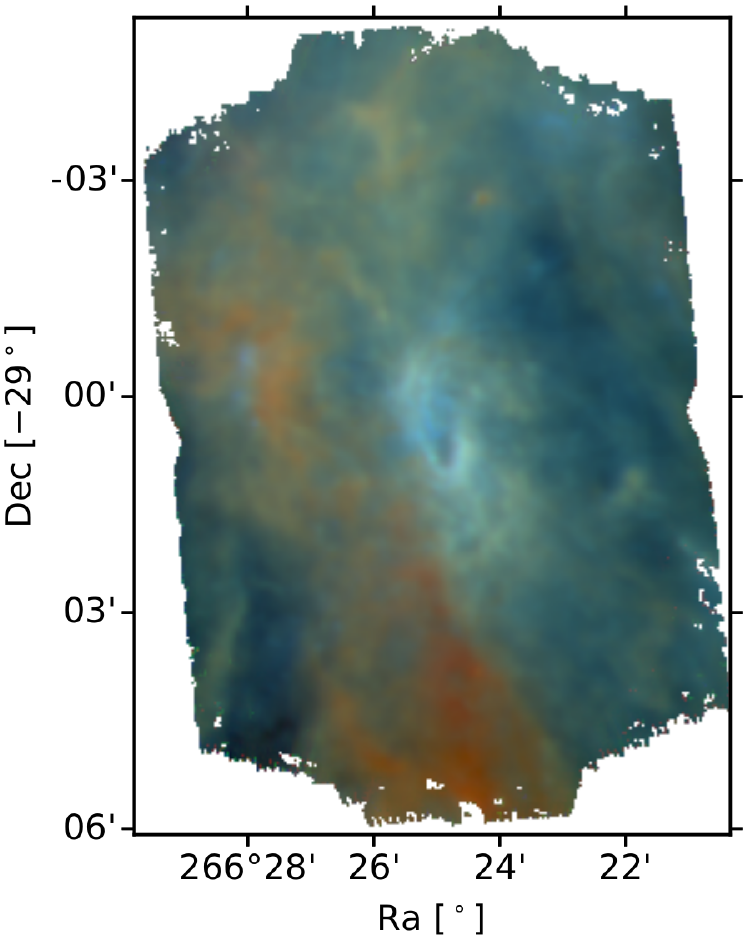}
	\captionof{figure}[RGB Image of the galactic center]{Composite FIR image of the Galactic Center, generated using the algorithm of \cite{2004PASP..116..133L}. We have scaled the intensity of the red band according to $I'_r = I_r^{0.9}$, the intensity of the green band $I'_g =I_g^{0.6}$ and the intensity of the blue band $I'_b =I_b^{0.5}$.}
	\label{rgbImage}
\end{Figure}

To remove these artefacts, we computed the offsets and aligned each cube with its first map. Naively, one would expect that this removes the pointing offset errors. The pointing errors, however, impair the performance of JScanam. This is because the pointing errors in the individual exposures smear out the averaged image of all individual exposures of an observation. This averaged image is used by JScanam to robustly calculate the detector read-out noise. Therefore, the pointing errors hinder an optimal removal of the detector read out noise, which in turn leads to an imprecise calculation of the offsets. 

To overcome this, the pointing correction needs to be handled iteratively. In example, we need to re-reduce the pointing-corrected cube and re-compute the pointing offsets several times until we end up with the final pointing offset corrected data cube. We refer to Appendix \ref{pointingCorrection} for details of this procedure. A similar procedure has been applied by \cite{2016ApJ...825...32S} for their Herschel/SPIRE maps. Our procedure creates a pointing-corrected data cube of $40$ images per observation. We plot a color composite image obtained in this manner in Figure \ref{rgbImage}. The image shown is the highest resolution images of the Galactic Center in the FIR to date. The median images of the individual bands are shown in Appendix \ref{medianMaps}.

\subsubsection{Median subtraction and affine coordinate transform}
\label{MedianSubtraction}
Next, we perform a pixel-wise median subtraction. In order to align the maps with the median map and to correct for other linear distortions, we apply affine coordinate transformations to the individual maps. The parameters are obtained numerically from minimizing the residual maps.
This produces a data cube of $40$ residual maps for each observation.

\subsubsection{Periodic pattern removal}
In the residual maps, a periodic strip pattern is the dominant artefact. 

To remove this pattern, we Fourier transformed the residual maps. In the Fourier transformed maps, the periodic pattern manifests itself as a few symmetrical peaks. We masked these peaks with the median intensity of the Fourier-transformed maps and back transformed the masked maps.

\subsubsection{Linear drift removal} 
\label{linearFitting}
For each cube, we noticed a small linear drift of the flux, i.e. the residual maps showed a linear increase or decrease of flux over the course of each observation. We verified that this is the case for pixels at least one beam away from Sgr~A*. This trend can be removed, pixel by pixel, by subtracting a linear fit from each pixel's light curve. Or, in more technical terms, we remove the linear trend by fitting and subtracting a linear function along the time axis of the residual map data cube for each spatial pixel.

\subsubsection{Smoothing and running mean}
\label{resmoothing}
In order to smooth any remaining smaller-than-resolution artefacts, we convolved the residual maps with the band's respective PSF, which is available from the instrument control center's (ICC) website\footnote{\url{http://tinyurl.com/pacs-psf}}. We corrected the PSF for the missing energy fraction as provided by the ICC and adjusted the pixel scale.

In addition to the spatial smoothing, we computed a temporal running mean for each map of width three. 

\subsubsection{Manual fine tuning}
\label{additionalRefinement}
The median subtraction (Subsec. \ref{MedianSubtraction}) and the linear drift removal procedure (Subsec. \ref{linearFitting}) assume that there is no source flux. Variable flux from Sgr~A* will appear as an excursion in the light curves of the respective pixels, effectively skewing our linear drift correction. This issue can be overcome in the case when the increase or decrease of flux from Sgr~A* happens only for a part of the observation. In this case, we reiterate steps \ref{MedianSubtraction} to \ref{resmoothing}, excluding images and maps with excess flux at the position of Sgr A*.

However, such a procedure requires a priori knowledge of the presence of flux and potential outliers can be mistaken as flux from Sgr~A*. In consequence, we only apply this manual fine tuning of the reduction in the case when a believable flux excursion is detected (i.e., when the bands are correlated or a point source is discernible in the residual maps). Once we opted for such a manual fine tuning, we applied it to all pixels of a map equally. 

Explicitly, we applied this manual fine tuning to the observations of March 15, 17 and 19. The details of the manual fine tuning are discussed in Appendix \ref{ARoLC}.

\subsection{Light curves}
\label{lightCurveSubsec}
In order to obtain light curves of Sgr~A* we calculated the best fit amplitude $C$ of the ICC PSF to the pixel in which Sgr~A* is expected to be found. We weighted the fit with the standard deviation maps provided by the standard reduction. As the maps were smoothed with the PSF (Subsection \ref{resmoothing}), we smoothed the PSF with itself. This accounts for the wider FWHM of point sources in the smoothed map. The FWHM ($\hat{\sigma}$) of a Gauss fit, fitted to the convolved PSF, yields: $\mathbf{\hat{\sigma}_{r}}=(15.7",19.0")$; $\mathbf{\hat{\sigma}_{g}}=(9.0",10.0")$ and $\mathbf{\hat{\sigma}_{b}}=(8.7",9.7")$. The Gauss fit is allowed to rotate.

\subsubsection{Error}
\label{reference_error}
Because of the complicated source structure at the Galactic Centre, we decided to use reference regions as a proxy to estimate the photometric noise, as the formal fit error would not capture the true uncertainty. The reference regions were chosen by applying the following selection criteria:

a) The median intensity of the pixel in question should lie within $0.3$ to $2$ times the intensity of the Sgr~A* pixel in the median image.

b) The pixel in question should lie within $44$ pixels ($\widehat{=}~66"$, $74.8"$ and $123.2"$ for the blue, green and red band respectively) of Sgr~A*.

c) All pixels within one beam of Sgr~A* are excluded as reference points.

These constraints ensure that: 

a) only regions of the sky are chosen which have a comparable intensity (and therefore photon noise) to that of Sgr~A*;

b) enough scanning coverage\footnote{The scanning coverage corresponds to the ratio of the exposure time of an actual camera pixel and that of an image pixel. Due to pixelation this is not constant and degrades quickly at the borders of the image. This results in a higher uncertainty for pixels with low scanning coverage, for details see drizzle method \citealt{2002PASP..114..144F} and HIPE documentation.} is guaranteed, and the coverage is approximately constant;

c) the variability of Sgr~A* does not perturb the estimate of the noise.

To calculate the noise, we draw $40$ uncorrelated random positions within the reference regions. We then extracted light curves of the reference points. The scatter of the these reference light curves serves as a proxy to the noise. In the figures below, the reference light curves are represented by thin grey lines.

We compute the error on $C$ as the sum of the spatial and temporal variance:

\begin{itemize}
\item We calculate the standard deviation $SD_{t_n}(x,y)$ of the reference light curves for each map at time $t_n$. $SD_{t_n}(x,y)$ probes the quality of the reduction for each map.

\end{itemize}

\begin{itemize}
\item In addition, we calculated the mean of the standard deviation $SD_{ref}$ of the reference light curves. The mean of $SD_{ref}$ measures the intrinsic variation of the maps. 
\end{itemize}
We estimate the error of Sgr~A*'s flux as the quadratic sum of these two values:
\begin{equation}
\sigma_{t_n} = \sqrt{<SD_{ref}>^2+SD_{t_n}(x,y)^2}
\label{error}
\end{equation}
where $\sigma_{t_n}$ is the error for each map. 

The temporal error $SD_{ref}$ and the spatial error $SD_{t_n}(x,y)$ are correlated. Our ansatz overproduces the real error and thus is a conservative estimate of the error.

\end{multicols}
\begin{Figure}
	\centering
		\includegraphics[width=\textwidth,keepaspectratio,trim={3.5cm 0cm 1.5cm 0cm},clip]{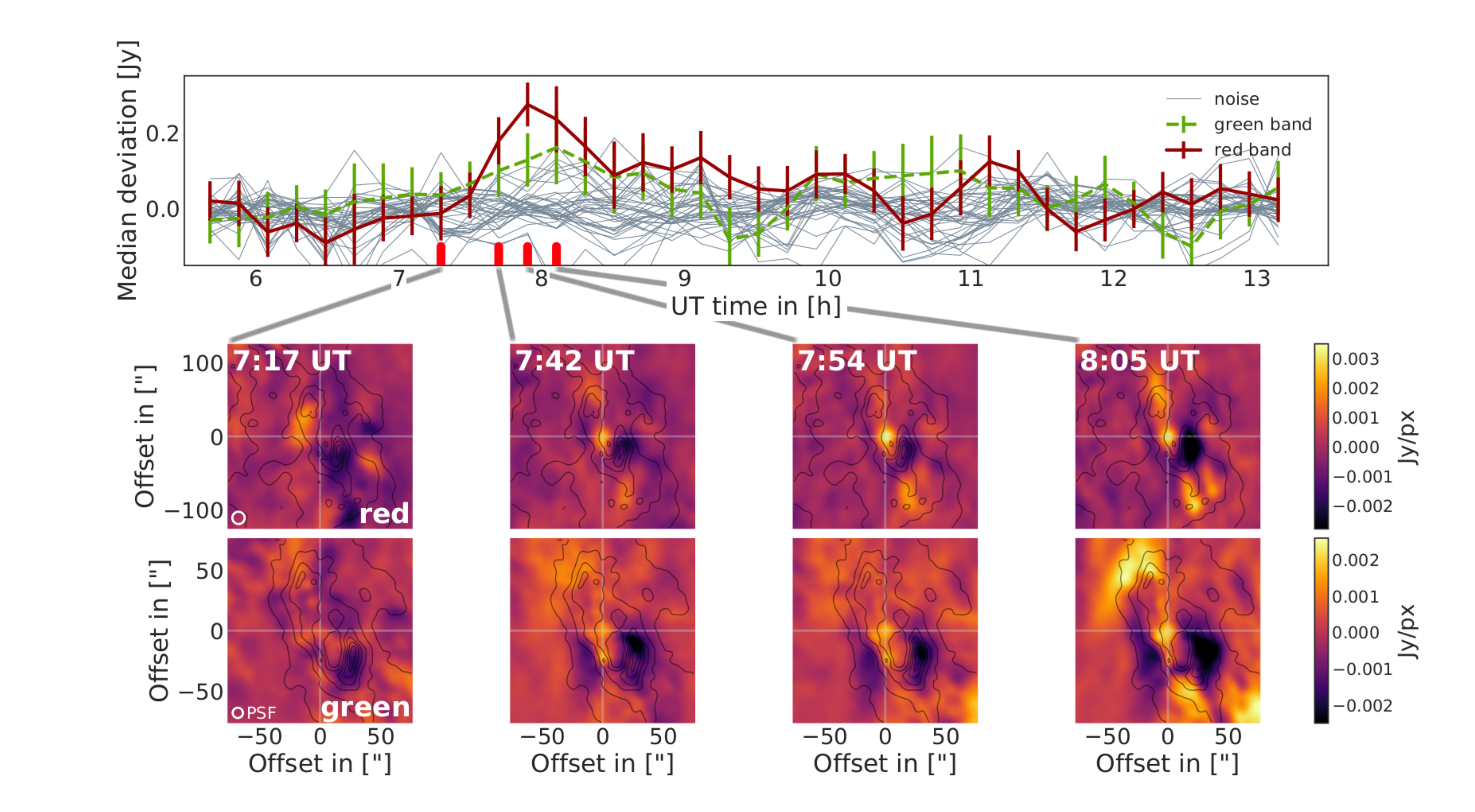}

	\captionof{figure}[The FIR variability of Sgr~A* on March 17]{The FIR variability on March 17: The upper panel shows the light curves of the red and green bands, as well as the reference light curves of the red band. Below are the residual maps which show the variable flux of Sgr~A*. The contour lines are intensity profiles of the respective median images. A point source is visible  at the position of Sgr~A*.}
	\label{2Dmarch17}
\end{Figure}
\begin{multicols}{2}

\subsection{Parallel observations}
The parallel NIR observations were obtained with NACO \citep{2003SPIE.4841..944L,2003SPIE.4839..140R}, and the images were reduced  following the procedure described in \cite{2009ApJ...698..676D}. We aligned the images using the bright isolated star S30. We combined images without discernible flares and created a median image. This median image was then subtracted from the individual images, creating residual maps. Aperture photometry was performed on the residual maps and the standard deviation of region without apparent sources between S2 and S30 calculated. We calibrate the flux as the ratio to the median S2 flux, where we assume a flux of $17.1 ~\mathrm{mJy}$ in the K band and a blackbody \citep{2017ApJ...837...30G}.

The parallel X-ray observations are presented in \cite{2015MNRAS.454.1525P}. For the XMM-Newton observations the diffuse background emission dominates the the quiescence X-ray flux of Sgr~A*. 

To account for this we subtract the mean flux of all XMM-Newton observations from the light curves. The error of the background subtraction is estimated from the standard deviation of the light curves.

\section{Results}
\label{R}
For clarity we only discuss the March 17 and March 19 observations, for which we detect flux from Sgr~A*. The other observations are discussed in Appendix \ref{otherNights}. 

\end{multicols}
\begin{Figure}
	\centering
	\includegraphics[width=1\textwidth,trim={0.5cm 0 4cm 0},clip]{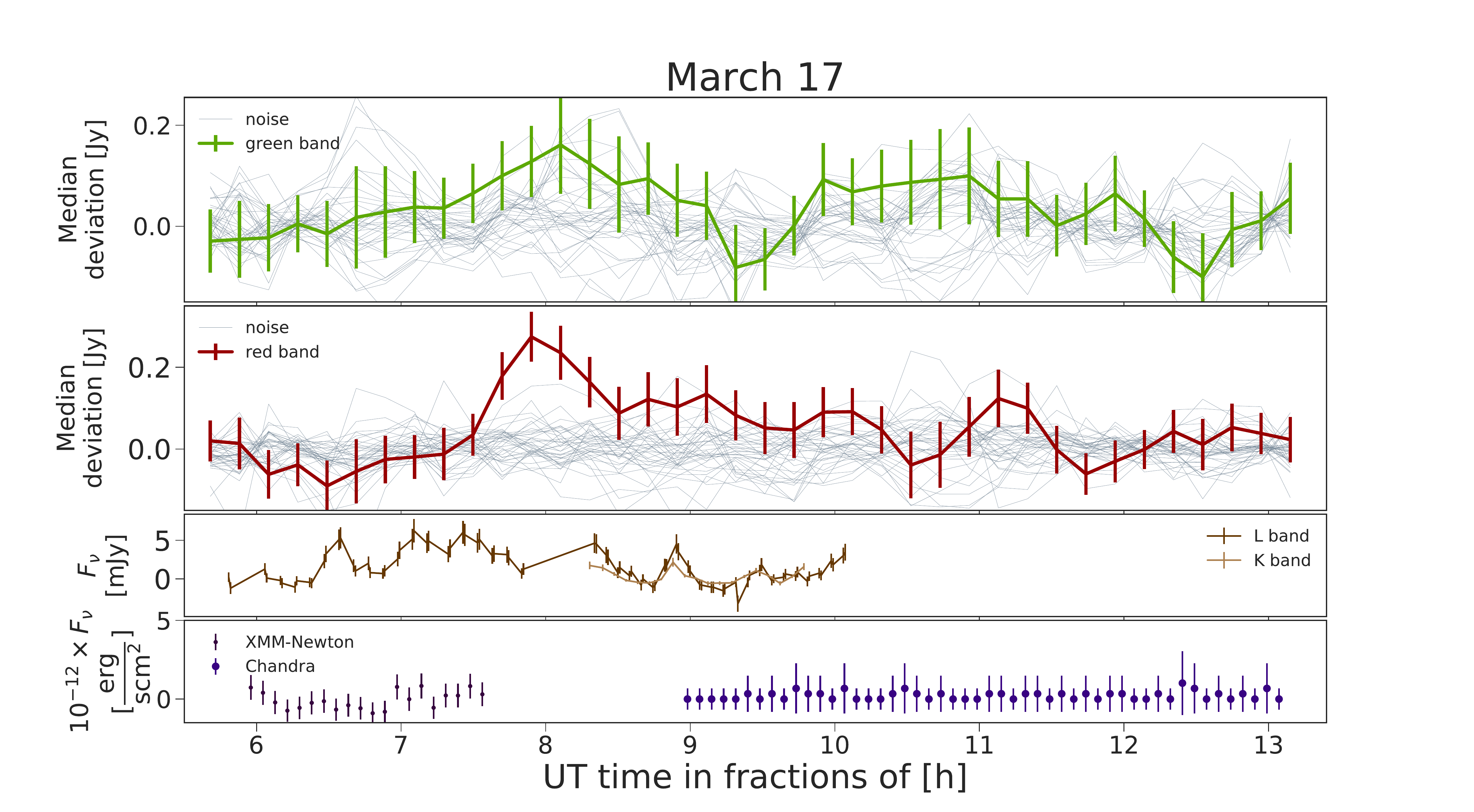}
	
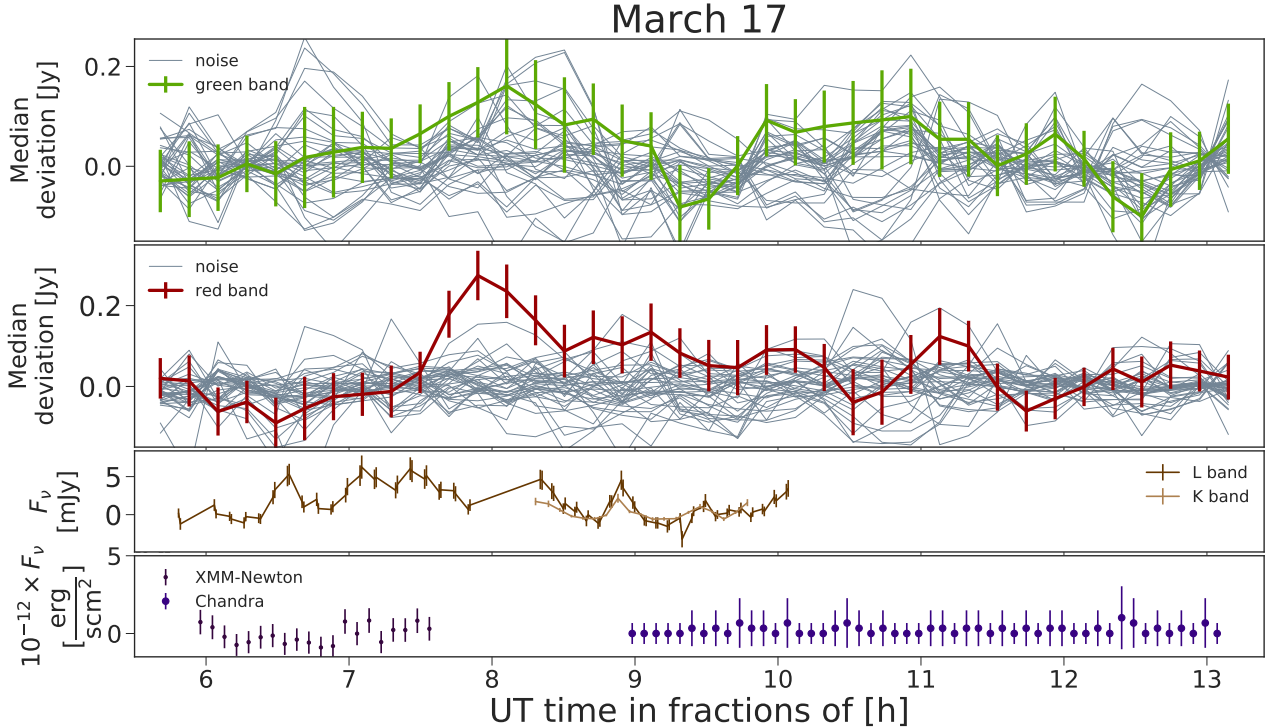
\captionof{figure}[Multiwavelength observation from March 17, 2012]{Multiwavelength observation from March 17, 2012. The top two panels give the red and green band FIR light curves. The grey lines are the light curves of the reference points. Below are the parallel K and L band NIR and $2-10~\mathrm{keV}$ X-ray observations.}
	\label{March17}
\end{Figure}
\begin{multicols}{2}

\subsection{Light curves}
\label{lightcurves}
\subsubsection{March 17}
\label{flare17}
Figure \ref{2Dmarch17} shows the light curves from the observations on March 17. A significant and correlated increase of flux was measured in both the red and the green band.

Defining the significance as the ratio of the peak flux to the error estimated from the reference light curves, the red band signal is significant at $4.5~\sigma$ and the green band is significant at $1.6~\sigma$. The flux peaks at around 8:20 UT to 8:30 UT. The red light curve remains above zero for about two hours. The green light curve drops to zero about an hour after the peak. 

Figure \ref{March17} shows all available light curves from this observation.

\paragraph{Comparison with the parallel observations}
The FIR activity is accompanied by NIR flaring with five consecutive, distinguishable peaks. There is no parallel X-ray flare. The first recorded NIR peak occurs roughly at 6:30 UT to 6:40 UT, which would imply a delay of $\sim 80~\mathrm{min}$ compared to the FIR peak. 
The association between the two events is unclear.

\subsubsection{March 19}
\label{flare19}
The light curves of the March 19 observation are shown in figure \ref{March19}. Since the flux appears to increase at the end of the light curve, the linear drift correction is less certain for this observation. In consequence, we do not use this observation to constrain the SED. However, the observation enhances the credibility of the detection on March 17, as the green and red FIR light curves again show a correlated increase towards the end of the observation (after 11 UT). Our best attempt at correcting the linear drifts yields a significance of $1.3\sigma$ for the red band and a significance of  $0.8\sigma$ for the green band.

\end{multicols}
\begin{Figure}
	\centering
	\includegraphics[width=1\textwidth,trim={0.5cm 0 4cm 0},clip]{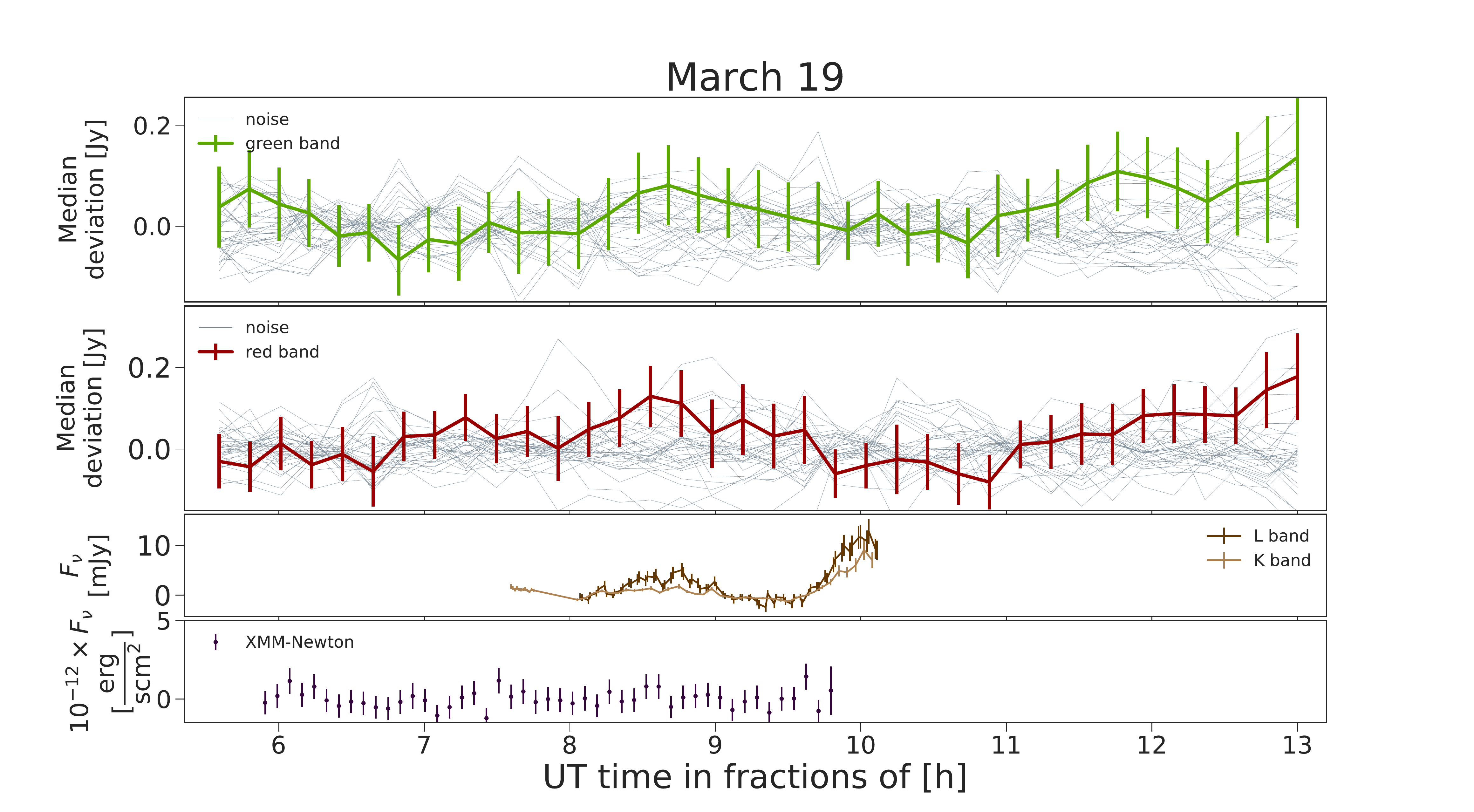}
	\captionof{figure}[Multiwavelength observation during March 19, 2012]{Multiwavelength observation during March 19, 2012. Same as figure \ref{March17} for the March 19 observation.}
	\label{March19}
\end{Figure}
\begin{multicols}{2}
\subparagraph{Comparison with parallel observations}
The first bump in the FIR light curve happens at 8:30 UT, during a NIR flare of intensity $\sim 14~\mathrm{Jy}$. Because of the low formal significance of $1.5 \sigma$, we cannot claim a detection here.  Unfortunately, there are no NIR parallel observations during the increase of flux after 11 UT.

However, it is interesting that there is a bright NIR flare at around 10 UT, without an immediate FIR counter part. This hints towards that the dominant variability process cannot be a simple extension of the NIR flares. Nevertheless, this bright NIR flare occurs about an hour to two hours earlier than the onset of the FIR activity. During our observing interval there is no X-ray flare. Unfortunately there are no parallel X-ray observations for the end of the observation. 

\subsection{Integrated residual maps}
\label{meanResi}
To increase statistics we sum the residual maps of each observation. The sum of the residual maps should only contain random fluctuations unless there is variable source in them, i.e. Sgr~A*.

For the March 17 observation and the red band, we find a point source located at the position of Sgr~A* (Figure~\ref{meanResidualAll}). We also find a point source in the corresponding integrated green band residual map.

\end{multicols}
\begin{Figure}
   \centering
  \includegraphics[width=1\textwidth,trim={0cm 0cm 0cm 0cm},clip]{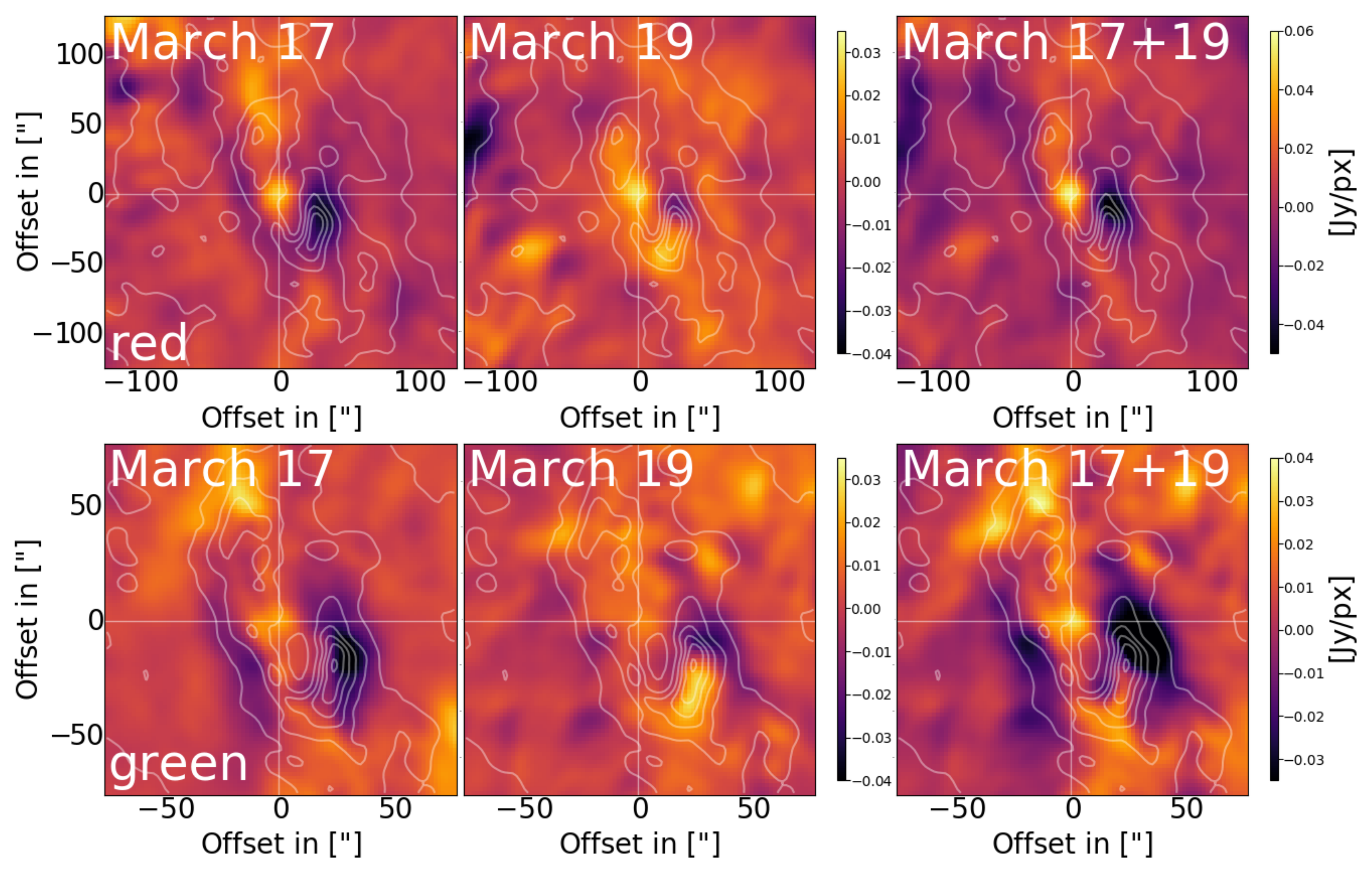} 
   \captionof{figure}[Integrated residual maps]{Integrated residual maps for the observations on March 17, March 19 and both nights combined. The left column shows the red and green integrated residual maps of the March 17 observation, the middle column the red and green integrated residual maps for March 19 and the right column shows the integrated residual maps of both nights.}
   \label{meanResidualAll}
\end{Figure}
\begin{multicols}{2}
The same is true for the March 19 observation and the red band integrated residual map: a point source is discernible at the location of Sgr~A*. The green excess is not strong enough to show up as a discernible point source in the corresponding integrated residual map. 

The integrated residual maps show large extended patches of positive and negative flux. These are spatially correlated with regions of high median intensity (and therefore not reference regions), as can be seen by comparing the patches with the contour lines. We suspect that at high fluxes, JScanam's baseline subtraction algorithm is less robust.

However, especially in the red band integrated residual maps, these patches are of significantly different morphology from that of a point source (see Appendix \ref{integratedResidualMaps_discussion}). In the green band the signal from Sgr~A* is weaker and thus the point source less pronounced.

In addition, while both observations show extended patches, the maps are, except at the position of Sgr~A*, not correlated across the different observations. 

\subsubsection{False alarm rate}
\label{falseAlarmRate}
To estimate how significant our detection is, we determine the probability of measuring a signal by chance. 

\end{multicols}
\begin{Table}
\resizebox{\textwidth}{!}{
\centering
\begin{tabular}{|c||c|c|c|c|}
\hline
\multirow{4}{*}{\textbf{Constraints}} & \multicolumn{2}{c|}{\multirow{2}{*}{\textbf{March 17}}} & \multicolumn{2}{c|}{\multirow{2}{*}{\textbf{March 19}}} \\
                              &     \multicolumn{2}{c|}{}                     &\multicolumn{2}{c|}{}\\
\cline{2-5}
                               & \multirow{2}{*}{\# of False Positives} & \multirow{2}{*}{Probability of Real Detection} & \multirow{2}{*}{\# of False Positives} & \multirow{2}{*}{Probability of Real Detection}\\ 
                 & \multirow{2}{*}{}  &  \multirow{2}{*}{} & &        \\
\hline 
\hline 
\multirow{2}{*}{a $1~\sigma$ significant signal in the red band}&\multirow{2}{*}{$8'203$}&\multirow{2}{*}{$82.7\%~\widehat{=}~1.36\sigma$}&\multirow{2}{*}{$3'102$}&\multirow{2}{*}{$82.7\%~ \widehat{=}~1.36\sigma$}\\ 
                                                          &                      &                      &                &\\
\hline
\multirow{2}{*}{a $2~\sigma$ significant signal in the red band}&\multirow{2}{*}{$600$}&\multirow{2}{*}{$98.7\%~\widehat{=}~2.48\sigma$}&\multirow{2}{*}{$103$}&\multirow{2}{*}{$n.a.$}\\ 
                                                        &                      &                      &                &\\
\hline
\multirow{2}{*}{a $3.8~\sigma$ significant signal in the red band}&\multirow{2}{*}{$1$}&\multirow{2}{*}{ $99.9979\%~\widehat{=}~4.25\sigma$}&\multirow{2}{*}{$1$}&\multirow{2}{*}{$n.a.$}\\ 
                                                        &                      &                      &                &\\
\hline
\multicolumn{1}{|c||}{a $1.3~\sigma$ significant signal in the red band }&\multirow{2}{*}{$882$}&\multirow{2}{*}{ $98.1\%~\widehat{=}~2.3\sigma$}&\multirow{2}{*}{$79$}&\multirow{2}{*}{$99.8\%~\widehat{=}~3.1\sigma$}\\ 
           and a $0.8~\sigma$ significant signal in the green band                        &                      &                      &                &\\
\hline
\multicolumn{1}{|c||}{a $2~\sigma$ significant signal in the red band }&\multirow{2}{*}{$35$}&\multirow{2}{*}{ $99.93\%~\widehat{=}~3.39\sigma$}&\multirow{2}{*}{$135$}&\multirow{2}{*}{$n.a.$}\\ 
          and a $1.5~\sigma$ significant signal in the green band&                      &                      &                &\\
\hline
\multirow{2}{*}{Number of tested pixels:} & \multicolumn{2}{|c|}{\multirow{2}{*}{$47'462$}}& \multicolumn{2}{c|}{\multirow{2}{*}{$45'362$}}\\
                 & \multicolumn{2}{c|}{} & \multicolumn{2}{c|}{}                                           \\
\hline
\end{tabular} 
}
\captionof{table}{False positive rate computed using the March 17 and March 19 observations.}
\label{falsePositives}
\end{Table}
\begin{multicols}{2}

In order to compute the false alarm rate, we measure the amplitudes at all valid reference pixels. Since the pixel scale as well as the median images are different between the two bands, we have to choose common reference regions. We apply the same criteria as before but make sure they are met in both bands.
For the $38$ residual maps of size $100~ \mathrm{px} \times 100~\mathrm{px}$  ($=380'000~px$) and the March 17 observation we find $47462$ pixels which are valid reference pixels in both bands. 

We compare the measured amplitude for each reference pixel with the error as given by equation \ref{error} and compute a significance. We then count the number of reference pixels with amplitudes above a given significance threshold (Table \ref{falsePositives}) and compare this with our observations.

The peak of the red band observation is significant at $\sim4.5\sigma$. We find no equally significant false alarm. For a significance of $3.8 \sigma$, there is one equally significant false positive within the tested pixels of the March 17 observation. This translates into a probability of $>99.998\%$ of the detection being real. In addition we observe a simultaneous $1.6\sigma$ significant green peak.

Note that we have estimated the errors conservatively, as a sum of the spatial and temporal variance. A conservative error estimate results in fewer points that have SNR of greater than one. For this reason, our $1\sigma$ constraint yields a probability of $82.2\%$ of the detection being real, rather than the expected $\sim68\%$.

For the March 19 observation, accounting for the systematic errors as before, we find $79$ false positives that are $1.3~\sigma$ significant in the red band and $0.8~\sigma$ significant in the green band. This corresponds to a $99.8\%$ probability of the detection being real. The number of false positives is lower than for March 17. This reflects that our estimate of the systematic error is conservative.

\subsubsection{Summary of false alarm rate}
We have found no false positives that are as significant as the measured flux increase at the position of Sgr~A* for the March 17 observation. In addition:

\begin{itemize}
\item A point source is discernible at the proper location.
\item This point source can be found in two bands.
\item The flux is temporally correlated between the two bands.
\item While the green and red detector sit in the same instrument, they are independent from one another, probe different physical phenomena (warmer/colder dust) and the reductions are handled independently. 
\item There is a second observation on March 19, for which we can detect a correlated increase of flux.
\item When binning all maps together we find a discernible point source in two different observations, and two different bands. 
\item The residual maps for the different observations are not temporally correlated. The point source (Sgr~A*) is the only reoccurring spatial structure. 
	 
\end{itemize}
We conclude, therefore, that the measured increase in flux is due to a change in brightness of Sgr~A*.


\section{Discussion}
\label{discussion}

\subsection{Implications for the SED}
We now discuss these findings and compare the results to existing models of the accretion flow. The subtraction of the median in our maps precludes the possibility of absolute flux measurements. In consequence, our measurements are measurements of the variable flux components and are therefore lower limits on the total flux at the time of our measurement. 

In order to constrain the SED, we estimate a median flux based on the observed variable flux component. If we assume a fractional variability $r$, we can compute the constant component that was subtracted:
\begin{equation}
F_{\nu; median} = \frac{F_{\nu; variable}}{r}
\label{variabilityEq}
\end{equation}

Therefore, our detection together with a constraint on the fractional variability $r$ allows one to estimate the median flux.

\subsection{Constraining the variability}
The range of the fractional variability $r$ can be estimated either by comparing $r$ with the typical variability in other wavelength regimes or from theoretical arguments. 

When we assume a minimal fractional variability $r_{min}$, equation \ref{variabilityEq} turns into an equation for an upper limit of the flux. Thus, assuming that Sgr~A* is at least as variable as a certain value leads to upper limits.

Alternatively, it is possible to obtain a value for the fractional variability from theoretical predictions. Time-dependent simulations of accretion flows can in some cases yield a prediction for typical values of the variability. This prediction can consequently be used to obtain an estimate of the median flux.

\begin{Figure}
\centering
	\includegraphics[width=\textwidth,trim={0cm 0cm 1.75cm 0},clip]{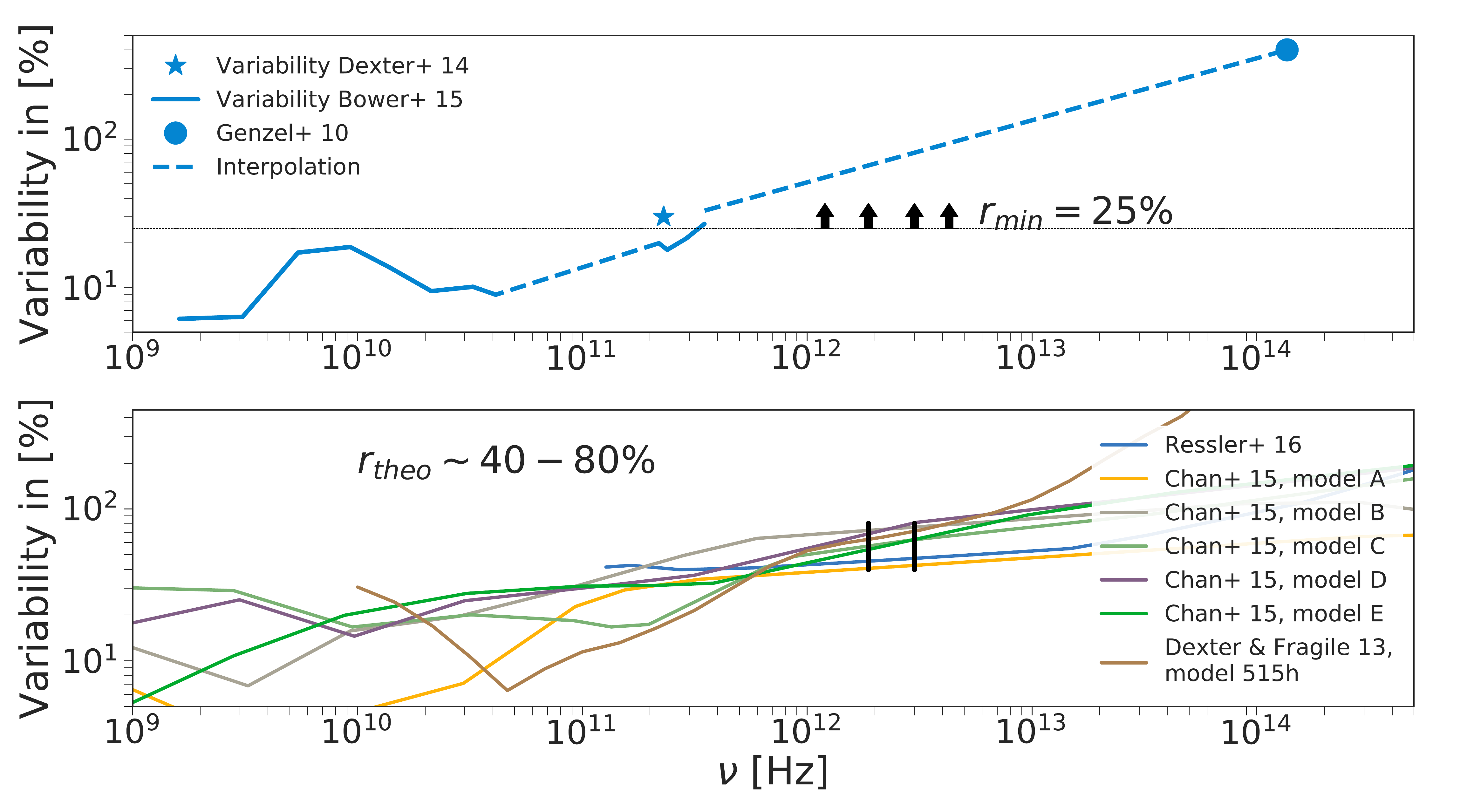}
	\captionof{figure}[Variability of Sgr~A* from observations and theory]{Variability of Sgr~A* from observations and theory. Left: measured variability from \cite{2015ApJ...802...69B} (as calculated from their SED), \cite{2014MNRAS.442.2797D} and \cite{2010RvMP...82.3121G}. We plot our assumption of a minimal variability of $r_{min}=25\%$ as black arrows. Right: Theoretical predictions of the variability from \cite{2013MNRAS.432.2252D}, \cite{2015ApJ...812..103C} and \cite{2016arXiv161109365R}, as calculated from their SED. The range of the FIR variability is $r_{theo}=40 - 80 \%$.}
	\label{variabilityPlot}
\end{Figure}

\subsubsection{Constraints based on observations}
In the following, we summarize the variability in the mm, sub-mm and the NIR regime and argue that a minimal variability of $r_{min}=25\%$ is a reasonable assumption.

Sgr~A* is highly variable around the sub-mm bump, with a characteristic time scale of around eight hours \citep{2014MNRAS.442.2797D}. In a comprehensive study of cm and mm light curves of Sgr~A*, \cite{2015ApJ...802...69B} calculated RMS variabilities from ALMA and VLA data. They find increasing RMS variabilities with decreasing wavelength and a variability around $30\%$ in the sub-mm. \cite{2014MNRAS.442.2797D} derived an RMS variability of $30\%$ at $1.3\mathrm{mm}$.

In the NIR, Sgr~A* is a highly variable source with regular faint flares and occasional bright flares. The brightest flares can easily exceed the faint flux by a factor of a few. \cite{2010RvMP...82.3121G} put the typical variability in the range of $300\%$ to $400\%$ and report a log-linear increase of variability throughout the spectrum. 
In both bands the variability is consistent with a red noise process (in the NIR e.g. \citealt{2009ApJ...691.1021D}, sub-mm e.g. \citealt{2014MNRAS.442.2797D}). This implies that the fractional variability depends on the time scale of the observation. 

The March 17 peak is the brightest in $40$ hours of observation. This is several times the typical variability time scale in the sub-mm. Since the variability time scale is  similar in the radio and mm regime (e.g. \citealt{2010RvMP...82.3121G}), it is reasonable to assume that the FIR variability time scale is not longer than the sub-mm one. Our observation length significantly exceeds this time scale and thus equation \ref{variabilityEq} estimates the median flux properly.  

Therefore, we assume that the minimal variability $r_{min}$ is at least as high as the long-term fractional variability observed in the sub-mm (Figure~\ref{variabilityPlot}).

\paragraph{Upper limits in the red and green band:}
Conservatively setting $r_{min}$ of the March 17 peak to $25\%$ we obtain:

$\langle F_{\nu \widehat{=} 160\mathrm{\mu m}}\rangle \leq (1.06\pm0.24) ~\mathrm{Jy}$ in the red band, and

$\langle F_{\nu \widehat{=} 100\mathrm{\mu m}}\rangle \leq (0.64\pm0.4)~\mathrm{Jy}$ in the green band. 

Because of the higher background in the green band, the uncertainty of the green band data is higher. In addition, the observation time was only $16$ hours, which makes applying eq. \ref{variabilityEq} less robust. 

We stress that these upper limits  would hold even if we had not detected Sgr~A*.

\paragraph{Upper limits for the blue band}
\label{blueBandUp}
We determine the standard deviation of the light curves of the reference pixels for the blue $70~\mathrm{\mu m}$ band. This is done for the March 15 and 21 observations.

The blue March 13 observation is impaired by a signal drift of unknown origin and therefore neglected. We use the blue band standard deviation of March 21 to compute the upper limit. The  $3\sigma$ limit for a non-detection is obtained by multiplying the standard deviation by a factor of three and dividing it by $0.25$ as before. This yields $\langle F_{\nu \widehat{=} 100\mathrm{\mu m}}\rangle \leq 0.84 ~ \mathrm{Jy}$ (see Appendix \ref{noiseCharacteristics} and \ref{otherNights} for details).

\subsubsection{Theoretical predictions for the FIR variability}
\label{theo_predict}
Several time-dependent simulations of the accretion flow of Sgr~A* exist which can reasonably reproduce the mm, sub-mm and/or NIR variability. As such they provide an estimate of the mean and the $1\sigma~\mathrm{RMS}$ variability. This gives a value for $r$, which we use to estimate the median flux. Examples of time-dependent simulations are \cite{2009ApJ...703L.142D}, \cite{2012ApJ...746L..10D}, \cite{2013MNRAS.432.2252D}, \cite{2015ApJ...812..103C} and \cite{2016arXiv161109365R}. 

We plot the variability prediction \cite{2013MNRAS.432.2252D}, \cite{2015ApJ...812..103C} and \cite{2016arXiv161109365R} in Figure \ref{variabilityPlot}.  The variability in these works ranges from $r_{theo}\sim 40\%$ to $r_{theo}\sim 80\%$. The mid range of these values is $r_{theo} \approx 60 \%$.
For the purpose of illustration, we choose this value as representative of current state of the art time-dependent simulations. Given the simplicity of equation \ref{variabilityEq}, it is straightforward to scale our results to find median flux densities corresponding to alternative values of $r_{theo}$.

Alternatively, time-dependent simulations can be directly tested against our observations. The variability prediction at  the FIR frequencies can be used to obtain $r_{theo}$ and the corresponding FIR median flux ad-hoc. This allows a self-consistent test of the parameters of any time-dependent simulation. Furthermore, if the flux distribution is known, the fact that we observe the brightest peak in $40$ hours can be used to estimate $r_{theo}$ even more accurately.

\end{multicols}
\begin{Figure}
\centering
	\includegraphics[width=\textwidth,trim={2.5cm 0cm 2.1cm 0},clip]{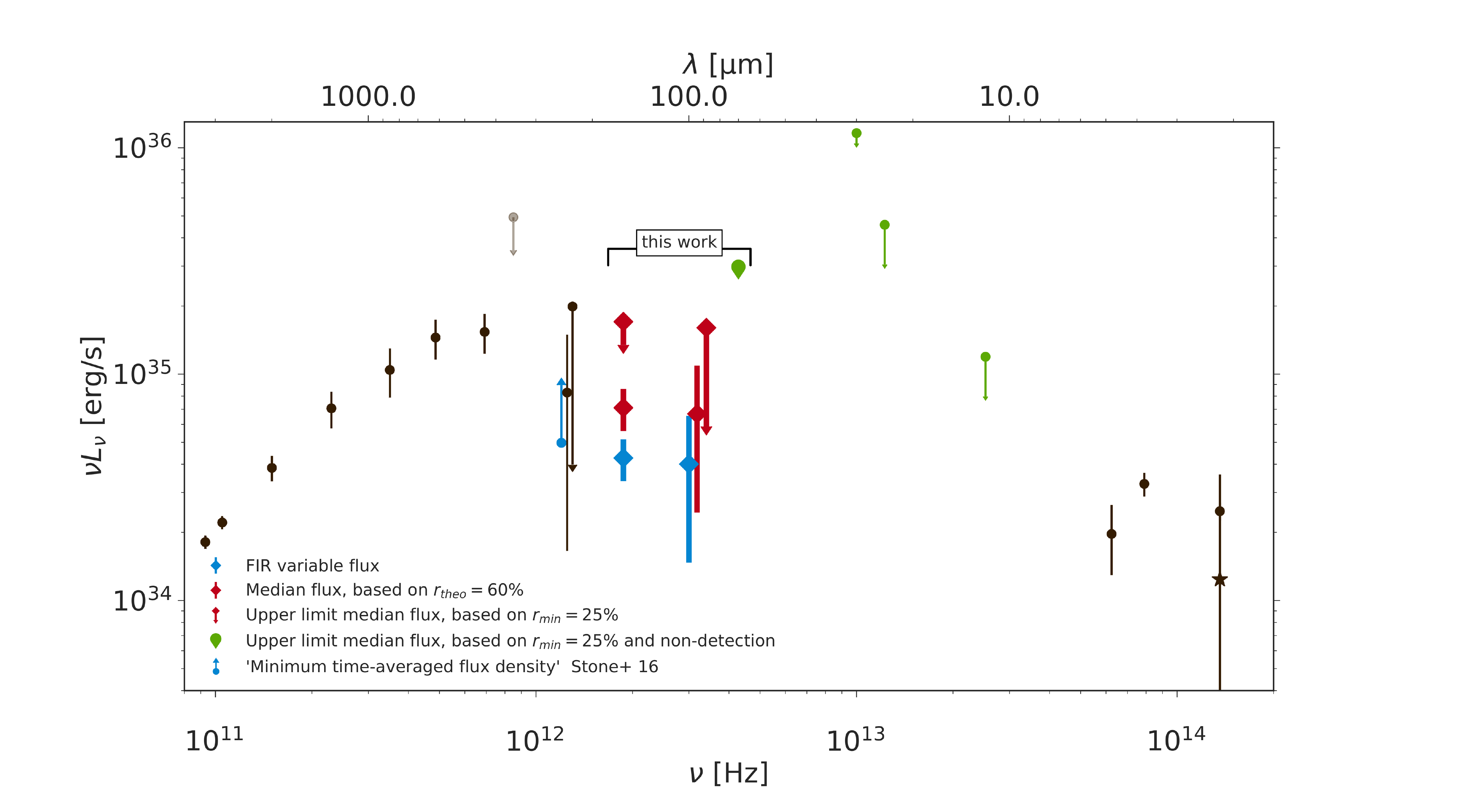}
	\captionof{figure}[Upper limits]{An updated SED of Sgr~A*:
	Measured mm to sub-mm data from left to right: \cite{2015AA...576A..41B}, \cite{1998ASPC..144..323F}, \cite{2015ApJ...802...69B}, \cite{2016AA...593A..44L}. At $\nu = 890~\mathrm{GHz}$, we show the measurement of \cite{1997ApJ...490L..77S} as an upper limit. This is because we believe that this measurement overestimates the flux due to exceptionally high flux at the time of the measurement. The blue point at $\nu = 1.2~\mathrm{THz}$ is the "minimum time-averaged flux density" of \cite{2016ApJ...825...32S}, where we have assigned an uncertainty of $0.4~\mathrm{Jy}$. 
Blue diamonds at $\nu = 1.9~\mathrm{THz}$ and $\nu = 3.0~\mathrm{THz}$ are our observed variable FIR flux. The upper limits at in the THz are based on our assumption of a minimal flux excursion of $25\%$. The data points below are the estimates of the median flux, based on a theoretical prediction of a $60\%$ fractional variability. The green upper limit at $\nu  = 4.3~\mathrm{THz}$ is based on the non-detection in the blue band. 
The MIR upper limits are taken from \cite{2001ARAA..39..309M}, \cite{2009ApJ...698..676D} and \cite{2011AA...532A..83S}.
In the NIR, the points denote mean fluxes measured by \cite{2011AA...532A..83S}, whereas the asterisk denotes the median reported by \cite{2010ApJ...725..450D}.
We plot values and constraints of the quiescence/median flux in dark brown, and the brighter flux excursions (e.g.\ our FIR measurements) in blue. Upper limits based on non-detections are plotted in green.}
	\label{SED_plain}	
\end{Figure} 
\begin{multicols}{2}

\paragraph{Theoretical prediction for the median flux}
Setting the variability to $r_{theo}=60\%$ we obtain:

$\langle F_{\nu \widehat{=} 160\mathrm{\mu m}}\rangle \approx 0.5\pm0.1~\mathrm{Jy}$ in the red band, and

$\langle F_{\nu \widehat{=} 100\mathrm{\mu m}}\rangle \approx 0.3\pm0.2~\mathrm{Jy}$ in the green band.

\subsection{An updated SED of Sgr A*}
In Figure \ref{SED_plain}, we plot our measurements of the FIR variable flux, the upper limits and the theoretical prediction of the median flux. 
For the cm, mm and sub-mm we use modern, high resolution data obtained from VLBI instruments such as ALMA and the VLA, where available.

\end{multicols}
\begin{Figure}
\includegraphics[width=\textwidth,trim={2.5cm 0cm 0cm 0},clip]{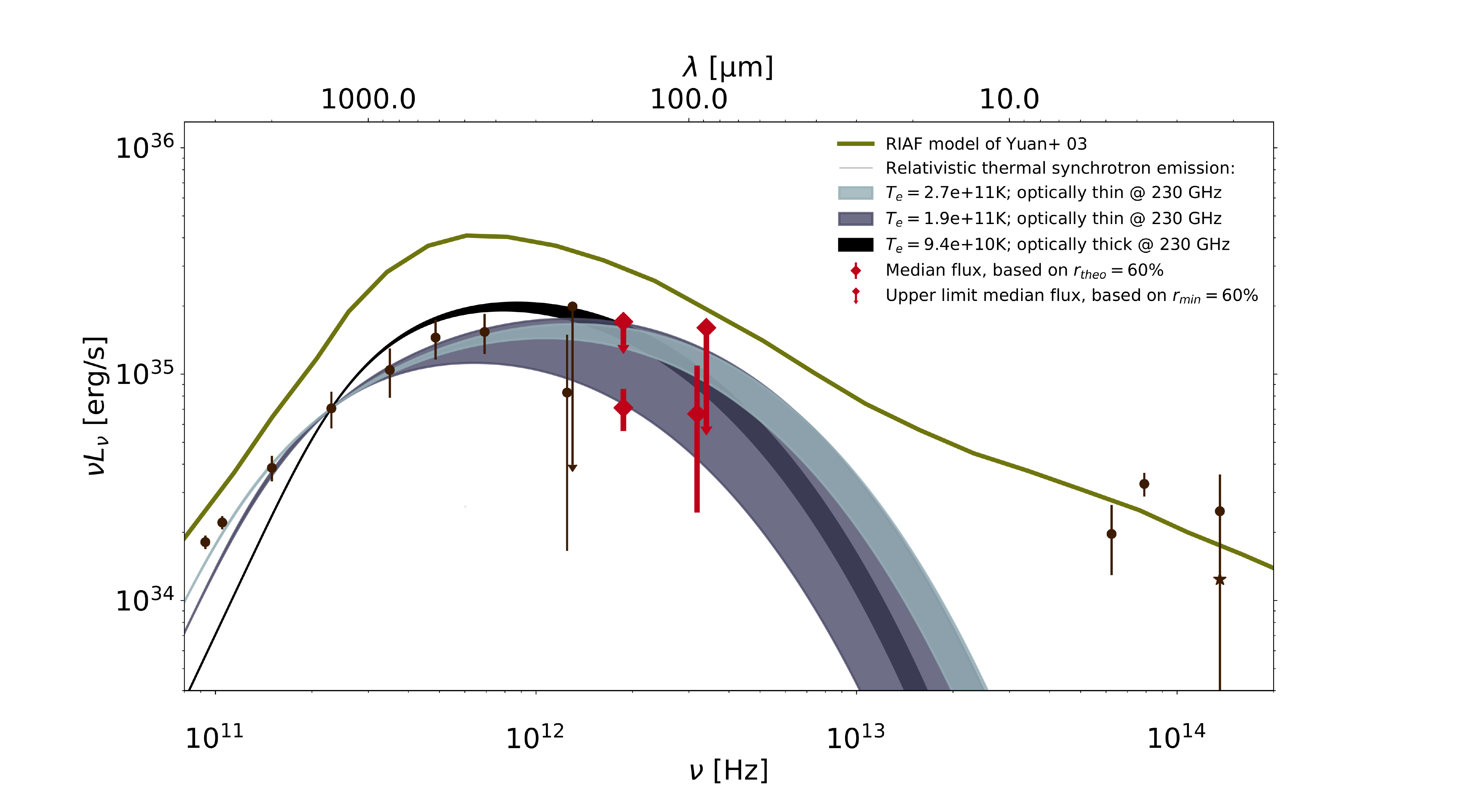}
\captionof{figure}[Upper limits]{RIAF Model for Sgr~A* compared to observations.\\
Same as Figure \ref{SED_plain}, without the FIR variable flux and MIR limits. Solid olive line is the 1D RIAF model of \cite{2003ANS...324..453Y}. The set of spectra below are synchrotron spectra of a  relativistic and thermal electron distribution. The width of the spectra demonstrate the slice through the parameter space of plasma-$\beta$ which are consistent with the observations. We show the spectra with the lowest electron temperature ($T_e=9.4\times 10^{10} ~\mathrm{K}$) that is consistent with our limits as well as sub-mm measurements. At $230~\mathrm{GHz}$ this spectrum is optically thick. The other two spectra shown are hotter and the plasma is optically thin. Here the peak is broad and set by $\nu / \nu_c \sim 1$ and not the optical depth. }

\label{SED_upperLimits}
\end{Figure}
\begin{multicols}{2}
\subsection{The "submillimeter bump" and spherical models of the accretion flow}

The model plotted in Figure \ref{SED_upperLimits} is the quiescence/median flux of the \cite{2003ANS...324..453Y} RIAF model. The original model overproduces the flux throughout much of the mm and sub-mm regime and is also inconsistent with our new FIR upper limits. In fact, our data as well as modern ALMA and VLA data show that the mm and sub-mm SED is less 'bumpy' than assumed in the original model (and older single dish observations, e.g. \citealt{1998ASPC..144..323F}). Therefore, the notion of a "sub-mm bump" may be outdated. 

In 1D RIAF models, the mm and sub-mm regime luminosity is dominated by emission from a spherical bulge of hot electrons with a thermal energy distribution. We approximate such a spherical bulge of hot electrons by assuming a thermal distribution of electrons in a region with radius $R$ with constant density, temperature, and magnetic field strength. The radius is set to be $R = 40~ \mathrm{\mu as}$, based on the mm-VLBI size \citep{2008Natur.455...78D}. Using the \textit{symphony} code of \cite{2016ApJ...822...34P} to compute the emission and absorption coefficients, we obtain the luminosity of such a configuration. We assume a wide range of values for the magnetic field strengths defined by the plasma parameter $\beta = 0.03 - 238$. To obtain the electron density, we normalize the flux to the observed value at $230~\mathrm{GHz}$. This yields a wide range of spectra from which we select the physically plausible ones and comparing them with the observed SED. We find that, a thermal distribution of electrons can describe the observed luminosity in the sub-mm and FIR regime and that the electron temperature is of the order of $T_e \sim 10^{11}~ \mathrm{K}$. 

Such calculations are rather sensitive to the radius of the hot bulge of electrons and the normalization flux assumed. Therefore, this electron temperature is only an estimate.

We proceed by computing the optical depth $\tau$ for our parameter grid. At $230~\mathrm{GHz}$, the accretion flow is optically thin for most valid solutions.  Only for two solutions, with $T_e<1.1\times10^{11}~\mathrm{K}$, is the optical depth $\tau$ greater than $1$. For the optically thin solutions, the peak is broad and the turn-over is set by $\nu  / \nu_c \sim 1$ and not the optical depth. 

This is interesting in the context of polarization measurements of Sgr A*. Synchrotron radiation from an optically thin, relativistic thermal distribution is expected to be highly polarized \citep{1979ApJ...228..268J}.
Faraday rotation, on the other hand, can scramble the polarization significantly, but is sensitive to both the optical depth and the electron temperature (e.g. \citealt{2016MNRAS.462..115D}). Models where the peak is set by synchrotron self-absorption are expected to be optically thick and depolarized by internal Faraday rotation. Higher temperatures, like the ones favored here, are more consistent with the $\sim 5-10\%$ linear polarization observed in Sgr A* \citep{2018MNRAS.tmp.1156J}.

In addition, we have also considered a power-law and a $\kappa$-distribution for the electron energy distribution. We find that a single power-law distribution with $\gamma_{min} \sim 350 - 500$ and $p \sim 3 - 4$ could explain both the sub-mm and the NIR emission (but not the radio spectrum). On the other hand, it is difficult to model the far- to near-infrared spectrum with the $\kappa$-distribution. For models that can successfully match the NIR median flux, the flux contribution from power-law electrons is too high.

\section{Summary and Outlook}
\label{SaO}
We have, for the first time, detected Sgr~A* in the far infrared. There are four immediate conclusions from this:

\begin{itemize}
	\item Sgr~A* is a variable source at $160~\mathrm{\mu m}$ and $100~\mathrm{\mu m}$. The observed peak deviation from median flux at $160~\mu m$ is $\Delta F_{\nu}=(0.27\pm0.07)~\mathrm{Jy}$ and at $100~\mathrm{\mu m}$ $\Delta F_{\nu}=(0.16\pm0.10)~\mathrm{Jy}$.
	\item The measured variability only places lower limits on the flux for the time of the measurement. Nevertheless, the measured peak variability can be used to constrain the SED by assuming a variability. Models with a prediction of the variability can be tested directly.
	\item Assuming a minimal flux excursion of $25\%$ over a period of $40$ hours allows us to compute upper limits in the red and green bands. At $160~\mathrm{\mu m}$ the upper limit is $\langle F_{\nu}\rangle \leq (1.06\pm0.24) ~\mathrm{Jy}$ and at $100~\mathrm{\mu m}$ the upper limit is $\langle F_{\nu}\rangle \leq (0.64\pm0.4) ~\mathrm{Jy}$. Using the 16 hours of non-detection in the blue band we compute a $70 \mathrm{\mu m}$ upper limit of $\langle F_{\nu}\rangle \leq 0.84 ~\mathrm{Jy}$.
	\item Theoretical predictions put the variability in the FIR in the range of $40 - 80 \%$. Using a theoretical variability of $\sim60\%$ yields an estimate for the FIR median flux of $\langle F_{\nu} \rangle \approx 0.5 \pm 0.1$ in the blue band and $\langle F_{\nu} \rangle \approx 0.3 \pm 0.2$ in the green band.
\end{itemize}

We find that modern VLA and ALMA data as well as our results show that the sub-mm flux of Sgr~A* is lower than in older observations. In consequence, we find that the 1D RIAF model by \cite{2003ANS...324..453Y}, which fitted the older sub-mm measurements well, is not consistent with the FIR upper limits and modern measurements of the sub-mm flux. In consequence, we argue that the overall shape of the sub-mm SED is less "bumpy" than previously assumed. 

Assuming an isotropic and spherical bulge of relativistic and thermally distributed electrons allows a simplistic implementation of an accretion flow model. Computing several plausible spectra of such a configuration reveals that our FIR measurements, as well as the modern ALMA and VLA data, can be described by such a configuration. The electron temperature is of the order of a few $10^{12}~\mathrm{K}$. This is slightly higher than older estimates. Computing the optical depth of the hot electron bulge, we find the electron plasma at $230~\mathrm{GHz}$ is optically thin for most valid solutions. For those solutions, the peak in the sub-mm is broad and the turn-over is set by $\nu/\nu_c\sim 1$ and not the optical depth.

\paragraph{Acknowledgment}
GP acknowledges support by the Bundesministerium fur Wirtschaft und Technologie/Deutsches Zentrum fur Luft- und Raumfahrt (BMWI/DLR, FKZ 50 OR 1408) and the Max Planck Society.
SvF, PP, IW, AJ-R and FW acknowledge support from the International  Max Planck Research School (IMPRS) on Astrophysics at the Ludwig-Maximilians University, as well as funding from the Max-Planck Society. 

\bibliographystyle{aa}
\bibliography{thesisBIB}

\begin{thebibliography}{64}
\expandafter\ifx\csname natexlab\endcsname\relax\def\natexlab#1{#1}\fi

\bibitem[{{Baganoff} {et~al.}(2001){Baganoff}, {Bautz}, {Ricker}, {Brandt},
  {Feigelson}, {Garmire}, {Townsley}, {Maeda}, \&
  {Morris}}]{2001AAS...199.8503B}
{Baganoff}, F.~K., {Bautz}, M.~W., {Ricker}, G.~R., {et~al.} 2001, in Bulletin
  of the American Astronomical Society, Vol.~33, American Astronomical Society
  Meeting Abstracts, 1429

\bibitem[{{Baganoff} {et~al.}(2003){Baganoff}, {Maeda}, {Morris}, {Bautz},
  {Brandt}, {Cui}, {Doty}, {Feigelson}, {Garmire}, {Pravdo}, {Ricker}, \&
  {Townsley}}]{2003ApJ...591..891B}
{Baganoff}, F.~K., {Maeda}, Y., {Morris}, M., {et~al.} 2003, \apj, 591, 891

\bibitem[{{Ball} {et~al.}(2016){Ball}, {{\"O}zel}, {Psaltis}, \&
  {Chan}}]{2016ApJ...826...77B}
{Ball}, D., {{\"O}zel}, F., {Psaltis}, D., \& {Chan}, C.-k. 2016, \apj, 826, 77

\bibitem[{{Barri{\`e}re} {et~al.}(2014){Barri{\`e}re}, {Tomsick}, {Baganoff},
  {Boggs}, {Christensen}, {Craig}, {Dexter}, {Grefenstette}, {Hailey},
  {Harrison}, {Madsen}, {Mori}, {Stern}, {Zhang}, {Zhang}, \&
  {Zoglauer}}]{2014ApJ...786...46B}
{Barri{\`e}re}, N.~M., {Tomsick}, J.~A., {Baganoff}, F.~K., {et~al.} 2014,
  \apj, 786, 46

\bibitem[{{Bower} {et~al.}(2015){Bower}, {Markoff}, {Dexter}, {Gurwell},
  {Moran}, {Brunthaler}, {Falcke}, {Fragile}, {Maitra}, {Marrone}, {Peck},
  {Rushton}, \& {Wright}}]{2015ApJ...802...69B}
{Bower}, G.~C., {Markoff}, S., {Dexter}, J., {et~al.} 2015, \apj, 802, 69

\bibitem[{{Brinkerink} {et~al.}(2015){Brinkerink}, {Falcke}, {Law}, {Barkats},
  {Bower}, {Brunthaler}, {Gammie}, {Impellizzeri}, {Markoff}, {Menten},
  {Moscibrodzka}, {Peck}, {Rushton}, {Schaaf}, \&
  {Wright}}]{2015AA...576A..41B}
{Brinkerink}, C.~D., {Falcke}, H., {Law}, C.~J., {et~al.} 2015, \aap, 576, A41

\bibitem[{{Broderick} \& {Narayan}(2006)}]{2006ApJ...638L..21B}
{Broderick}, A.~E. \& {Narayan}, R. 2006, \apjl, 638, L21

\bibitem[{{Chael} {et~al.}(2017){Chael}, {Narayan}, \& {Sa{\c
  d}owski}}]{2017MNRAS.470.2367C}
{Chael}, A.~A., {Narayan}, R., \& {Sa{\c d}owski}, A. 2017, \mnras, 470, 2367

\bibitem[{{Chan} {et~al.}(2015){Chan}, {Psaltis}, {{\"O}zel}, {Medeiros},
  {Marrone}, {Sa{\c d}owski}, \& {Narayan}}]{2015ApJ...812..103C}
{Chan}, C.-k., {Psaltis}, D., {{\"O}zel}, F., {et~al.} 2015, \apj, 812, 103

\bibitem[{{Dexter}(2016)}]{2016MNRAS.462..115D}
{Dexter}, J. 2016, \mnras, 462, 115

\bibitem[{{Dexter} {et~al.}(2009){Dexter}, {Agol}, \&
  {Fragile}}]{2009ApJ...703L.142D}
{Dexter}, J., {Agol}, E., \& {Fragile}, P.~C. 2009, \apjl, 703, L142

\bibitem[{{Dexter} {et~al.}(2010){Dexter}, {Agol}, {Fragile}, \&
  {McKinney}}]{2010ApJ...717.1092D}
{Dexter}, J., {Agol}, E., {Fragile}, P.~C., \& {McKinney}, J.~C. 2010, \apj,
  717, 1092

\bibitem[{{Dexter} \& {Fragile}(2013)}]{2013MNRAS.432.2252D}
{Dexter}, J. \& {Fragile}, P.~C. 2013, \mnras, 432, 2252

\bibitem[{{Dexter} {et~al.}(2014){Dexter}, {Kelly}, {Bower}, {Marrone},
  {Stone}, \& {Plambeck}}]{2014MNRAS.442.2797D}
{Dexter}, J., {Kelly}, B., {Bower}, G.~C., {et~al.} 2014, \mnras, 442, 2797

\bibitem[{{Diolaiti} {et~al.}(2000){Diolaiti}, {Bendinelli}, {Bonaccini},
  {Close}, {Currie}, \& {Parmeggiani}}]{2000SPIE.4007..879D}
{Diolaiti}, E., {Bendinelli}, O., {Bonaccini}, D., {et~al.} 2000, in \procspie,
  Vol. 4007, Adaptive Optical Systems Technology, ed. P.~L. {Wizinowich},
  879--888

\bibitem[{{Do} {et~al.}(2009){Do}, {Ghez}, {Morris}, {Yelda}, {Meyer}, {Lu},
  {Hornstein}, \& {Matthews}}]{2009ApJ...691.1021D}
{Do}, T., {Ghez}, A.~M., {Morris}, M.~R., {et~al.} 2009, \apj, 691, 1021

\bibitem[{{Dodds-Eden} {et~al.}(2009){Dodds-Eden}, {Porquet}, {Trap},
  {Quataert}, {Haubois}, {Gillessen}, {Grosso}, {Pantin}, {Falcke}, {Rouan},
  {Genzel}, {Hasinger}, {Goldwurm}, {Yusef-Zadeh}, {Clenet}, {Trippe},
  {Lagage}, {Bartko}, {Eisenhauer}, {Ott}, {Paumard}, {Perrin}, {Yuan},
  {Fritz}, \& {Mascetti}}]{2009ApJ...698..676D}
{Dodds-Eden}, K., {Porquet}, D., {Trap}, G., {et~al.} 2009, \apj, 698, 676

\bibitem[{{Dodds-Eden} {et~al.}(2010){Dodds-Eden}, {Sharma}, {Quataert},
  {Genzel}, {Gillessen}, {Eisenhauer}, \& {Porquet}}]{2010ApJ...725..450D}
{Dodds-Eden}, K., {Sharma}, P., {Quataert}, E., {et~al.} 2010, \apj, 725, 450

\bibitem[{{Doeleman} {et~al.}(2008){Doeleman}, {Weintroub}, {Rogers},
  {Plambeck}, {Freund}, {Tilanus}, {Friberg}, {Ziurys}, {Moran}, {Corey},
  {Young}, {Smythe}, {Titus}, {Marrone}, {Cappallo}, {Bock}, {Bower},
  {Chamberlin}, {Davis}, {Krichbaum}, {Lamb}, {Maness}, {Niell}, {Roy},
  {Strittmatter}, {Werthimer}, {Whitney}, \& {Woody}}]{2008Natur.455...78D}
{Doeleman}, S.~S., {Weintroub}, J., {Rogers}, A.~E.~E., {et~al.} 2008, \nat,
  455, 78

\bibitem[{{Dolence} {et~al.}(2012){Dolence}, {Gammie}, {Shiokawa}, \&
  {Noble}}]{2012ApJ...746L..10D}
{Dolence}, J.~C., {Gammie}, C.~F., {Shiokawa}, H., \& {Noble}, S.~C. 2012,
  \apjl, 746, L10

\bibitem[{{Etxaluze} {et~al.}(2011){Etxaluze}, {Smith}, {Tolls}, {Stark}, \&
  {Gonz{\'a}lez-Alfonso}}]{2011AJ....142..134E}
{Etxaluze}, M., {Smith}, H.~A., {Tolls}, V., {Stark}, A.~A., \&
  {Gonz{\'a}lez-Alfonso}, E. 2011, \aj, 142, 134

\bibitem[{{Falcke} {et~al.}(1998){Falcke}, {Goss}, {Ho}, {Matsuo}, {Teuben},
  {Wilson}, {Zhao}, \& {Zylka}}]{1998ASPC..144..323F}
{Falcke}, H., {Goss}, W.~M., {Ho}, L.~C., {et~al.} 1998, in Astronomical
  Society of the Pacific Conference Series, Vol. 144, IAU Colloq. 164: Radio
  Emission from Galactic and Extragalactic Compact Sources, ed. J.~A. {Zensus},
  G.~B. {Taylor}, \& J.~M. {Wrobel}, 323

\bibitem[{{Falcke} \& {Markoff}(2000)}]{2000A&A...362..113F}
{Falcke}, H. \& {Markoff}, S. 2000, \aap, 362, 113

\bibitem[{{Fruchter} \& {Hook}(2002)}]{2002PASP..114..144F}
{Fruchter}, A.~S. \& {Hook}, R.~N. 2002, \pasp, 114, 144

\bibitem[{{Genzel} {et~al.}(2010){Genzel}, {Eisenhauer}, \&
  {Gillessen}}]{2010RvMP...82.3121G}
{Genzel}, R., {Eisenhauer}, F., \& {Gillessen}, S. 2010, Reviews of Modern
  Physics, 82, 3121

\bibitem[{{Genzel} {et~al.}(2003){Genzel}, {Sch{\"o}del}, {Ott}, {Eckart},
  {Alexander}, {Lacombe}, {Rouan}, \& {Aschenbach}}]{2003Natur.425..934G}
{Genzel}, R., {Sch{\"o}del}, R., {Ott}, T., {et~al.} 2003, \nat, 425, 934

\bibitem[{{Gillessen} {et~al.}(2017){Gillessen}, {Plewa}, {Eisenhauer}, {Sari},
  {Waisberg}, {Habibi}, {Pfuhl}, {George}, {Dexter}, {von Fellenberg}, {Ott},
  \& {Genzel}}]{2017ApJ...837...30G}
{Gillessen}, S., {Plewa}, P.~M., {Eisenhauer}, F., {et~al.} 2017, \apj, 837, 30

\bibitem[{{Graci{\'a}-Carpio} {et~al.}(2015){Graci{\'a}-Carpio}, {Wetzstein},
  \& {Roussel}}]{2015arXiv151203252G}
{Graci{\'a}-Carpio}, J., {Wetzstein}, M., \& {Roussel}, H. 2015, ArXiv e-prints

\bibitem[{{Howes}(2010)}]{2010MNRAS.409L.104H}
{Howes}, G.~G. 2010, \mnras, 409, L104

\bibitem[{{Ichimaru}(1977)}]{1977ApJ...214..840I}
{Ichimaru}, S. 1977, \apj, 214, 840

\bibitem[{{Jansen} {et~al.}(2001){Jansen}, {Lumb}, {Altieri}, {Clavel}, {Ehle},
  {Erd}, {Gabriel}, {Guainazzi}, {Gondoin}, {Much}, {Munoz}, {Santos},
  {Schartel}, {Texier}, \& {Vacanti}}]{2001A&A...365L...1J}
{Jansen}, F., {Lumb}, D., {Altieri}, B., {et~al.} 2001, \aap, 365, L1

\bibitem[{{Jim{\'e}nez-Rosales} \& {Dexter}(2018)}]{2018MNRAS.tmp.1156J}
{Jim{\'e}nez-Rosales}, A. \& {Dexter}, J. 2018, \mnras

\bibitem[{{Jones} \& {Hardee}(1979)}]{1979ApJ...228..268J}
{Jones}, T.~W. \& {Hardee}, P.~E. 1979, \apj, 228, 268

\bibitem[{{Lenzen} {et~al.}(2003){Lenzen}, {Hartung}, {Brandner}, {Finger},
  {Hubin}, {Lacombe}, {Lagrange}, {Lehnert}, {Moorwood}, \&
  {Mouillet}}]{2003SPIE.4841..944L}
{Lenzen}, R., {Hartung}, M., {Brandner}, W., {et~al.} 2003, in \procspie, Vol.
  4841, Instrument Design and Performance for Optical/Infrared Ground-based
  Telescopes, ed. M.~{Iye} \& A.~F.~M. {Moorwood}, 944--952

\bibitem[{{Liu} {et~al.}(2016){Liu}, {Wright}, {Zhao}, {Mills},
  {Requena-Torres}, {Matsushita}, {Mart{\'{\i}}n}, {Ott}, {Morris}, {Longmore},
  {Brinkerink}, \& {Falcke}}]{2016AA...593A..44L}
{Liu}, H.~B., {Wright}, M.~C.~H., {Zhao}, J.-H., {et~al.} 2016, \aap, 593, A44

\bibitem[{{Lupton} {et~al.}(2004){Lupton}, {Blanton}, {Fekete}, {Hogg},
  {O'Mullane}, {Szalay}, \& {Wherry}}]{2004PASP..116..133L}
{Lupton}, R., {Blanton}, M.~R., {Fekete}, G., {et~al.} 2004, \pasp, 116, 133

\bibitem[{{Mao} {et~al.}(2017){Mao}, {Dexter}, \&
  {Quataert}}]{2017MNRAS.466.4307M}
{Mao}, S.~A., {Dexter}, J., \& {Quataert}, E. 2017, \mnras, 466, 4307

\bibitem[{{Melia} \& {Falcke}(2001)}]{2001ARAA..39..309M}
{Melia}, F. \& {Falcke}, H. 2001, \araa, 39, 309

\bibitem[{{Mo{\'s}cibrodzka} \& {Falcke}(2013)}]{2013A&A...559L...3M}
{Mo{\'s}cibrodzka}, M. \& {Falcke}, H. 2013, \aap, 559, L3

\bibitem[{{Mo{\'s}cibrodzka} {et~al.}(2009){Mo{\'s}cibrodzka}, {Gammie},
  {Dolence}, {Shiokawa}, \& {Leung}}]{2009ApJ...706..497M}
{Mo{\'s}cibrodzka}, M., {Gammie}, C.~F., {Dolence}, J.~C., {Shiokawa}, H., \&
  {Leung}, P.~K. 2009, \apj, 706, 497

\bibitem[{{Narayan} \& {Yi}(1994)}]{1994ApJ...428L..13N}
{Narayan}, R. \& {Yi}, I. 1994, \apjl, 428, L13

\bibitem[{{Ott}(2010)}]{2010ASPC..434..139O}
{Ott}, S. 2010, in Astronomical Society of the Pacific Conference Series, Vol.
  434, Astronomical Data Analysis Software and Systems XIX, ed. Y.~{Mizumoto},
  K.-I. {Morita}, \& M.~{Ohishi}, 139

\bibitem[{{{\"O}zel} {et~al.}(2000){{\"O}zel}, {Psaltis}, \&
  {Narayan}}]{2000ApJ...541..234O}
{{\"O}zel}, F., {Psaltis}, D., \& {Narayan}, R. 2000, \apj, 541, 234

\bibitem[{{Pandya} {et~al.}(2016){Pandya}, {Zhang}, {Chandra}, \&
  {Gammie}}]{2016ApJ...822...34P}
{Pandya}, A., {Zhang}, Z., {Chandra}, M., \& {Gammie}, C.~F. 2016, \apj, 822,
  34

\bibitem[{{Pilbratt} {et~al.}(2010){Pilbratt}, {Riedinger}, {Passvogel},
  {Crone}, {Doyle}, {Gageur}, {Heras}, {Jewell}, {Metcalfe}, {Ott}, \&
  {Schmidt}}]{2010A&A...518L...1P}
{Pilbratt}, G.~L., {Riedinger}, J.~R., {Passvogel}, T., {et~al.} 2010, \aap,
  518, L1

\bibitem[{{Poglitsch} {et~al.}(2008){Poglitsch}, {Waelkens}, {Bauer}, {Cepa},
  {Feuchtgruber}, {Henning}, {van Hoof}, {Kerschbaum}, {Krause}, {Renotte},
  {Rodriguez}, {Saraceno}, \& {Vandenbussche}}]{2008SPIE.7010E..05P}
{Poglitsch}, A., {Waelkens}, C., {Bauer}, O.~H., {et~al.} 2008, in \procspie,
  Vol. 7010, Space Telescopes and Instrumentation 2008: Optical, Infrared, and
  Millimeter, 701005

\bibitem[{{Ponti} {et~al.}(2015){Ponti}, {De Marco}, {Morris}, {Merloni},
  {Mu{\~n}oz-Darias}, {Clavel}, {Haggard}, {Zhang}, {Nandra}, {Gillessen},
  {Mori}, {Neilsen}, {Rea}, {Degenaar}, {Terrier}, \&
  {Goldwurm}}]{2015MNRAS.454.1525P}
{Ponti}, G., {De Marco}, B., {Morris}, M.~R., {et~al.} 2015, \mnras, 454, 1525

\bibitem[{{Rees} {et~al.}(1982){Rees}, {Begelman}, {Blandford}, \&
  {Phinney}}]{1982Natur.295...17R}
{Rees}, M.~J., {Begelman}, M.~C., {Blandford}, R.~D., \& {Phinney}, E.~S. 1982,
  \nat, 295, 17

\bibitem[{{Ressler} {et~al.}(2016){Ressler}, {Tchekhovskoy}, {Quataert}, \&
  {Gammie}}]{2016arXiv161109365R}
{Ressler}, S.~M., {Tchekhovskoy}, A., {Quataert}, E., \& {Gammie}, C.~F. 2016,
  ArXiv e-prints

\bibitem[{{Rousset} {et~al.}(2003){Rousset}, {Lacombe}, {Puget}, {Hubin},
  {Gendron}, {Fusco}, {Arsenault}, {Charton}, {Feautrier}, {Gigan}, {Kern},
  {Lagrange}, {Madec}, {Mouillet}, {Rabaud}, {Rabou}, {Stadler}, \&
  {Zins}}]{2003SPIE.4839..140R}
{Rousset}, G., {Lacombe}, F., {Puget}, P., {et~al.} 2003, in \procspie, Vol.
  4839, Adaptive Optical System Technologies II, ed. P.~L. {Wizinowich} \&
  D.~{Bonaccini}, 140--149

\bibitem[{{S{\'a}nchez-Portal} {et~al.}(2014){S{\'a}nchez-Portal}, {Marston},
  {Altieri}, {Aussel}, {Feuchtgruber}, {Klaas}, {Linz}, {Lutz}, {Mer{\'{\i}}n},
  {M{\"u}ller}, {Nielbock}, {Oort}, {Pilbratt}, {Schmidt}, {Stephenson}, \&
  {Tuttlebee}}]{2014ExA....37..453S}
{S{\'a}nchez-Portal}, M., {Marston}, A., {Altieri}, B., {et~al.} 2014,
  Experimental Astronomy, 37, 453

\bibitem[{{Sch{\"o}del} {et~al.}(2011){Sch{\"o}del}, {Morris}, {Muzic},
  {Alberdi}, {Meyer}, {Eckart}, \& {Gezari}}]{2011AA...532A..83S}
{Sch{\"o}del}, R., {Morris}, M.~R., {Muzic}, K., {et~al.} 2011, \aap, 532, A83

\bibitem[{{Serabyn} {et~al.}(1997){Serabyn}, {Carlstrom}, {Lay}, {Lis},
  {Hunter}, {Lacy}, \& {Hills}}]{1997ApJ...490L..77S}
{Serabyn}, E., {Carlstrom}, J., {Lay}, O., {et~al.} 1997, \apjl, 490, L77

\bibitem[{{Shcherbakov} {et~al.}(2012){Shcherbakov}, {Penna}, \&
  {McKinney}}]{2012ApJ...755..133S}
{Shcherbakov}, R.~V., {Penna}, R.~F., \& {McKinney}, J.~C. 2012, \apj, 755, 133

\bibitem[{{Stone} {et~al.}(2016){Stone}, {Marrone}, {Dowell}, {Schulz},
  {Heinke}, \& {Yusef-Zadeh}}]{2016ApJ...825...32S}
{Stone}, J.~M., {Marrone}, D.~P., {Dowell}, C.~D., {et~al.} 2016, \apj, 825

\bibitem[{{Str{\"u}der} {et~al.}(2001){Str{\"u}der}, {Briel}, {Dennerl},
  {Hartmann}, {Kendziorra}, {Meidinger}, {Pfeffermann}, {Reppin}, {Aschenbach},
  {Bornemann}, {Br{\"a}uninger}, {Burkert}, {Elender}, {Freyberg}, {Haberl},
  {Hartner}, {Heuschmann}, {Hippmann}, {Kastelic}, {Kemmer}, {Kettenring},
  {Kink}, {Krause}, {M{\"u}ller}, {Oppitz}, {Pietsch}, {Popp}, {Predehl},
  {Read}, {Stephan}, {St{\"o}tter}, {Tr{\"u}mper}, {Holl}, {Kemmer}, {Soltau},
  {St{\"o}tter}, {Weber}, {Weichert}, {von Zanthier}, {Carathanassis}, {Lutz},
  {Richter}, {Solc}, {B{\"o}ttcher}, {Kuster}, {Staubert}, {Abbey}, {Holland},
  {Turner}, {Balasini}, {Bignami}, {La Palombara}, {Villa}, {Buttler},
  {Gianini}, {Lain{\'e}}, {Lumb}, \& {Dhez}}]{2001A&A...365L..18S}
{Str{\"u}der}, L., {Briel}, U., {Dennerl}, K., {et~al.} 2001, \aap, 365, L18

\bibitem[{{Weisskopf} {et~al.}(2000){Weisskopf}, {Tananbaum}, {Van Speybroeck},
  \& {O'Dell}}]{2000SPIE.4012....2W}
{Weisskopf}, M.~C., {Tananbaum}, H.~D., {Van Speybroeck}, L.~P., \& {O'Dell},
  S.~L. 2000, in \procspie, Vol. 4012, X-Ray Optics, Instruments, and Missions
  III, ed. J.~E. {Truemper} \& B.~{Aschenbach}, 2--16

\bibitem[{{Xu} {et~al.}(2006){Xu}, {Narayan}, {Quataert}, {Yuan}, \&
  {Baganoff}}]{2006ApJ...640..319X}
{Xu}, Y.-D., {Narayan}, R., {Quataert}, E., {Yuan}, F., \& {Baganoff}, F.~K.
  2006, \apj, 640, 319

\bibitem[{{Yuan} {et~al.}(2003{\natexlab{a}}){Yuan}, {Markoff}, \&
  {Falcke}}]{2003ANS...324..453Y}
{Yuan}, F., {Markoff}, S., \& {Falcke}, H. 2003{\natexlab{a}}, Astronomische
  Nachrichten Supplement, 324, 453

\bibitem[{{Yuan} \& {Narayan}(2014)}]{2014ARA&A..52..529Y}
{Yuan}, F. \& {Narayan}, R. 2014, \araa, 52, 529

\bibitem[{{Yuan} {et~al.}(2003{\natexlab{b}}){Yuan}, {Quataert}, \&
  {Narayan}}]{2003ApJ...598..301Y}
{Yuan}, F., {Quataert}, E., \& {Narayan}, R. 2003{\natexlab{b}}, \apj, 598, 301

\bibitem[{{Yuan} {et~al.}(2004){Yuan}, {Quataert}, \&
  {Narayan}}]{2004ApJ...606..894Y}
{Yuan}, F., {Quataert}, E., \& {Narayan}, R. 2004, \apj, 606, 894

\bibitem[{{Yusef-Zadeh} {et~al.}(2009){Yusef-Zadeh}, {Bushouse}, {Wardle},
  {Heinke}, {Roberts}, {Dowell}, {Brunthaler}, {Reid}, {Martin}, {Marrone},
  {Porquet}, {Grosso}, {Dodds-Eden}, {Bower}, {Wiesemeyer}, {Miyazaki}, {Pal},
  {Gillessen}, {Goldwurm}, {Trap}, \& {Maness}}]{2009ApJ...706..348Y}
{Yusef-Zadeh}, F., {Bushouse}, H., {Wardle}, M., {et~al.} 2009, \apj, 706, 348

\bibitem[{{Zhao} {et~al.}(2003){Zhao}, {Young}, {Herrnstein}, {Ho}, {Tsutsumi},
  {Lo}, {Goss}, \& {Bower}}]{2003ApJ...586L..29Z}
{Zhao}, J.-H., {Young}, K.~H., {Herrnstein}, R.~M., {et~al.} 2003, \apjl, 586,
  L29

\end{thebibliography}

\appendix
 \section{Median maps}
\label{medianMaps}

\end{multicols}
\begin{Figure}
\centering
	\includegraphics[width=\textwidth]{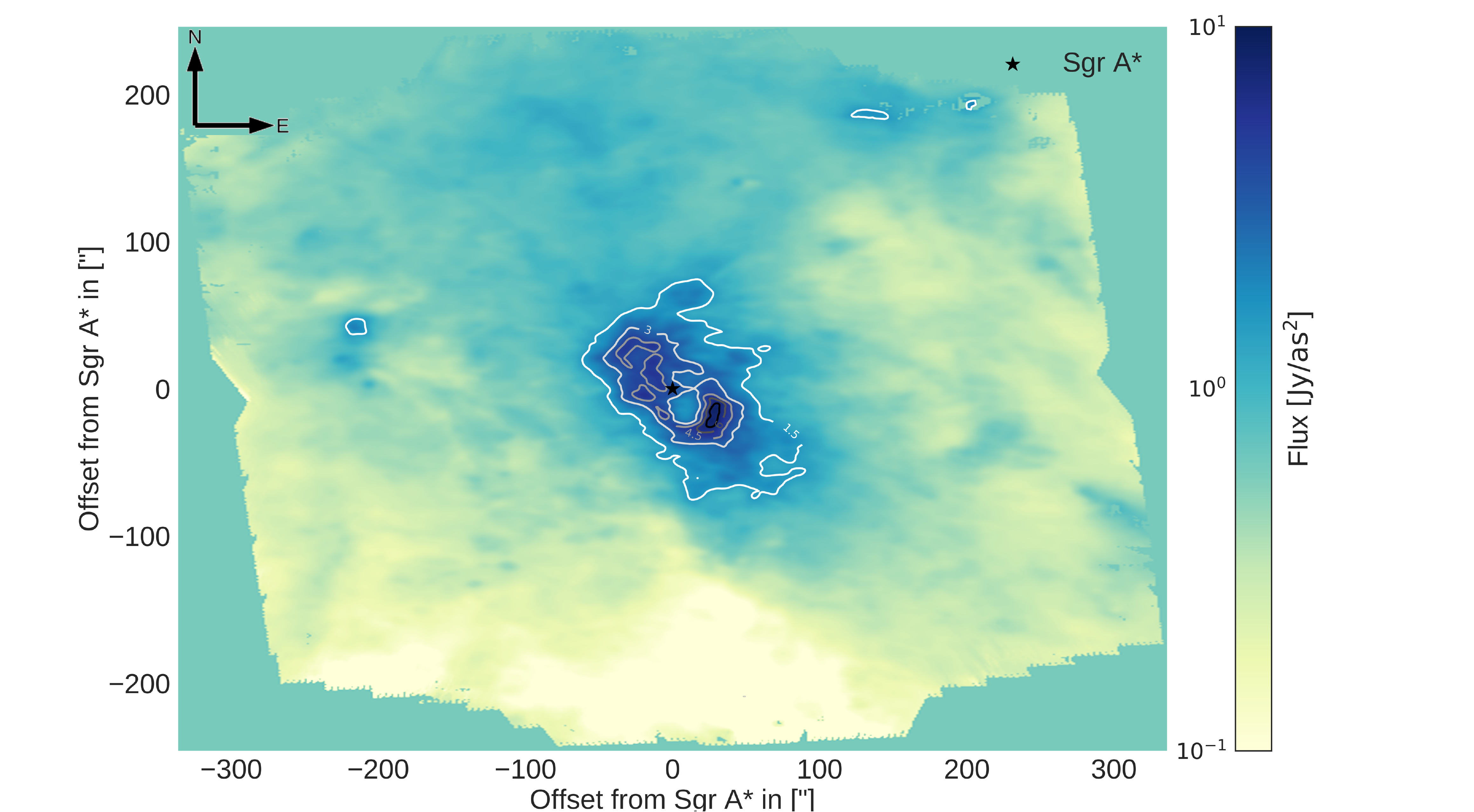}
	\captionof{figure}[Blue band median image]{Blue band median image of March 15 and 21. The integration time is $\sim 16$ hours. The color scale is logarithmic. JScanam creates images with relative intensities. To overcome this, we have normalized the images in Figure \ref{blueMedian}, \ref{greenMedian} and \ref{redMedian}, so that the pixel with the lowest flux value has a flux of $0 ~ \mathrm{Jy}$.}
	\label{blueMedian}
\end{Figure}
\begin{Figure}
\centering
	\includegraphics[width=\textwidth]{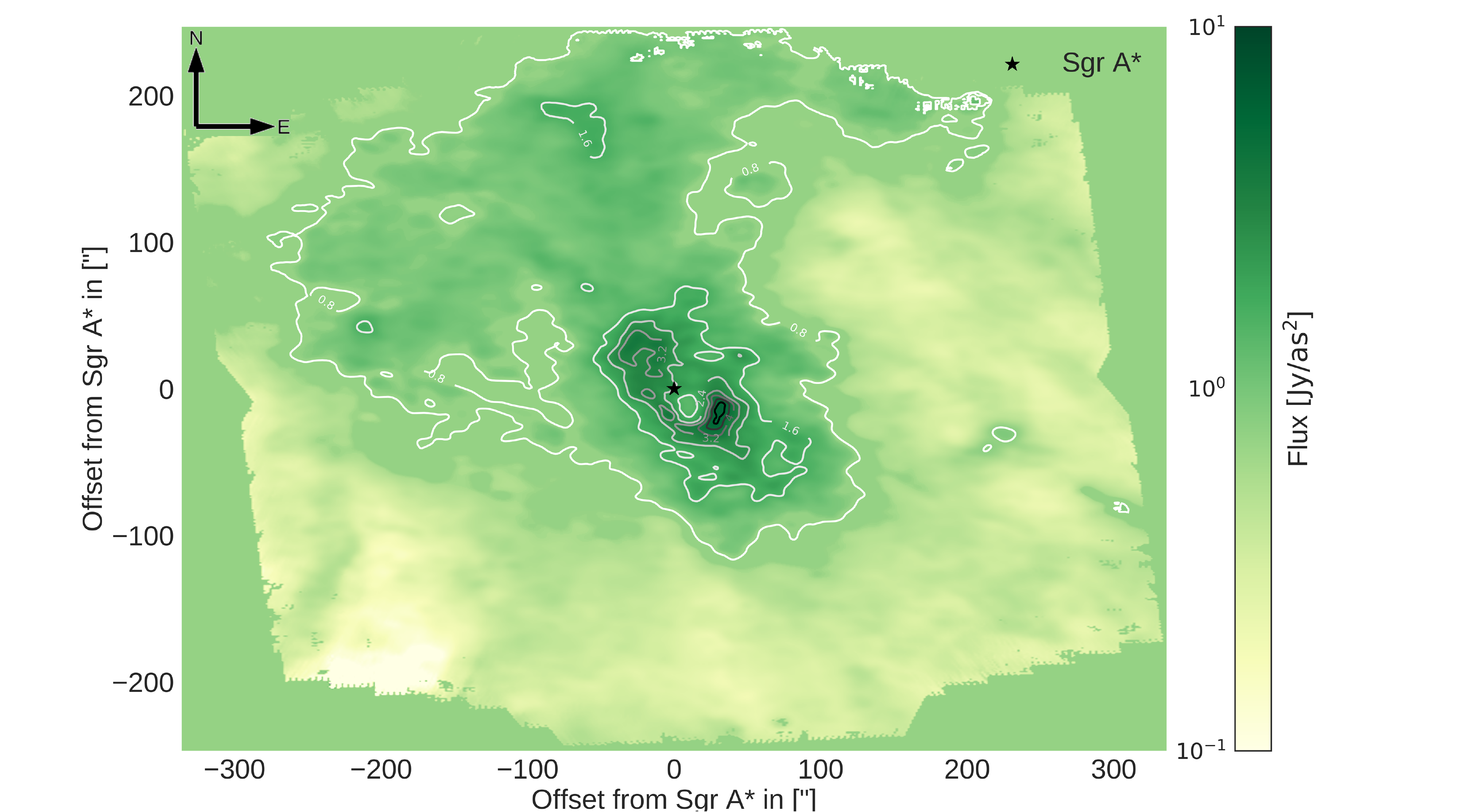}
	\captionof{figure}[Red band median image]{As in Figure \ref{blueMedian} for the green band median image. The observation dates are March 17 and 19, totaling to an integration time of $\sim 16$ hours.}
	\label{greenMedian}
\end{Figure}
\begin{Figure}
\centering
	\includegraphics[width=\textwidth]{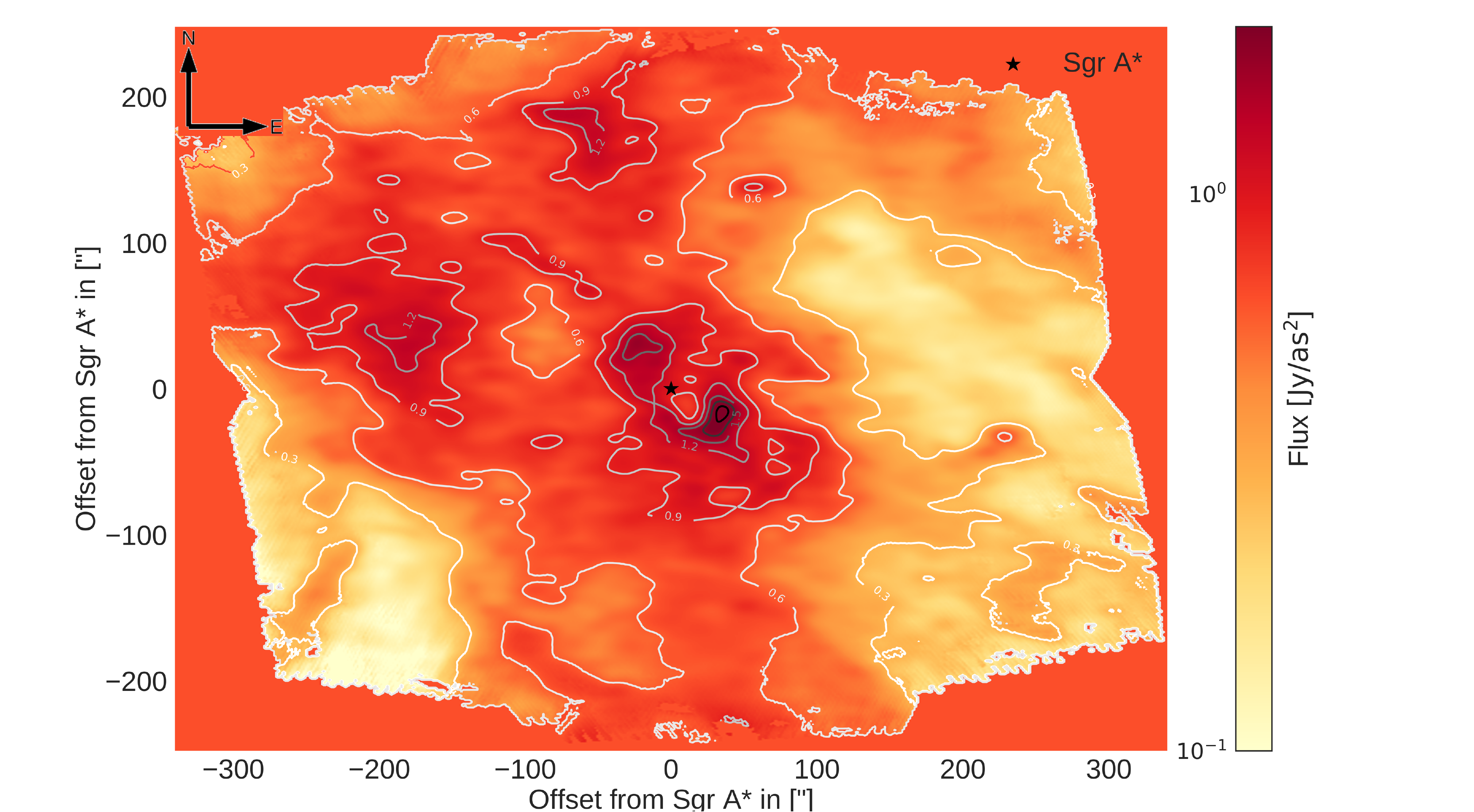}
	\captionof{figure}[Red band median image]{As in Figure \ref{blueMedian} for the red band median image and all nights. The integration time is around $40$ hours.}
	\label{redMedian}
\end{Figure}
\begin{multicols}{2}

We plot the median images of the three bands in Figures \ref{blueMedian}, \ref{greenMedian} and \ref{redMedian}. Since the images are pointing corrected, the images presented here are the highest resolution images of the Galactic Center to date.
Since JScanam does not produce images of absolute intensity we have normalized the maps such that the darkest pixel contains zero flux.

\section{Pointing offset correction}
\label{pointingCorrection}
Herschel experiences pointing offset errors. Simply aligning the images by shifting them on top of one another is not sufficient, as the pointing error smears out the images. This hinders the 1/f noise removal of JScanam from performing optimally. 
Therefore the pointing correction needs to be handled iteratively.

We correct the pointing offset as follows:
\begin{enumerate}
\item Reduce the raw level 2 data by running JScanam. For all observations, this creates sets of images impaired by the pointing errors. 
\item Compute the pointing offsets of these images using the HIPE method PhotHelper.getOptimalShift. This routine computes the offsets of the first image of an observation to the subsequent ones. The pointing offsets are then saved for further processing.
\item Correct the just calculated pointing offsets in the raw level 2 data, using the HIPE method PhotHelper.shiftFramesCoordinates\footnote{Both routines are available for use in HIPE version 15.0 2412}. This functions shifts the raw level 2 data, so that the offsets are neutralized. The shifted level 2 data of an observation is now, in first order, aligned to its first image. 
\item Rerun JScanam, using the shifted level 2 data. Since the images are now better aligned the averaged image of the observation is less smeared out. Because of that, 1/f noise removal of JScanam performs more efficiently. This allows for a sharper images, and therefore, when we recalculate the pointing offsets (repeat step 2) they decrease. 
\item Add the newly calculated pointing offsets (from step 3) together with the pointing offsets from the first iteration (step 2). The combined point offsets are now again applied to the raw level 2 data, shifting it. This creates a new set of shifted level 2 data.  
\item JScanam always uses two observations with scan directions for the reduction. These observations are tilted against each other and the scanning pattern is different. JScanam reduces both observations at the same time. Since both directions are impaired by the pointing offset error, the pointing offsets in one observation impair the calculation of the pointing offsets in the other observation. To minimize this effect, we restart the pointing offset correction from step 1. The difference to before is that we now always pair the raw level 2 data of one observation with the shifted level 2 data of another observation. The uncorrected observation is reduced together with the shifted one and its pointing offsets are determined and corrected as before (steps 1 to 5).
\item We iterate this last step four times, always determining the pointing offsets of one observation. After the last iteration, the pointing offsets in all observations are smaller than $0.05"$.
\end{enumerate}

\section{Noise characteristics}
\label{noiseCharacteristics}
We have verified that the fluxes in the reference pixels are approximately Gaussian distributed, see Figure \ref{histo}. This justifies the way we have calculated our error bars and the false alarm rate. Figure \ref{allHisto} shows the histograms for all nights. For March 19 the uncertainty in the manual fine tuning (c.f. \ref{additionalRefinement} and Appendix \ref{ARoLC}) causes a positive skew of the histogram. 
\begin{Figure}
\centering
	\includegraphics[width=\textwidth]{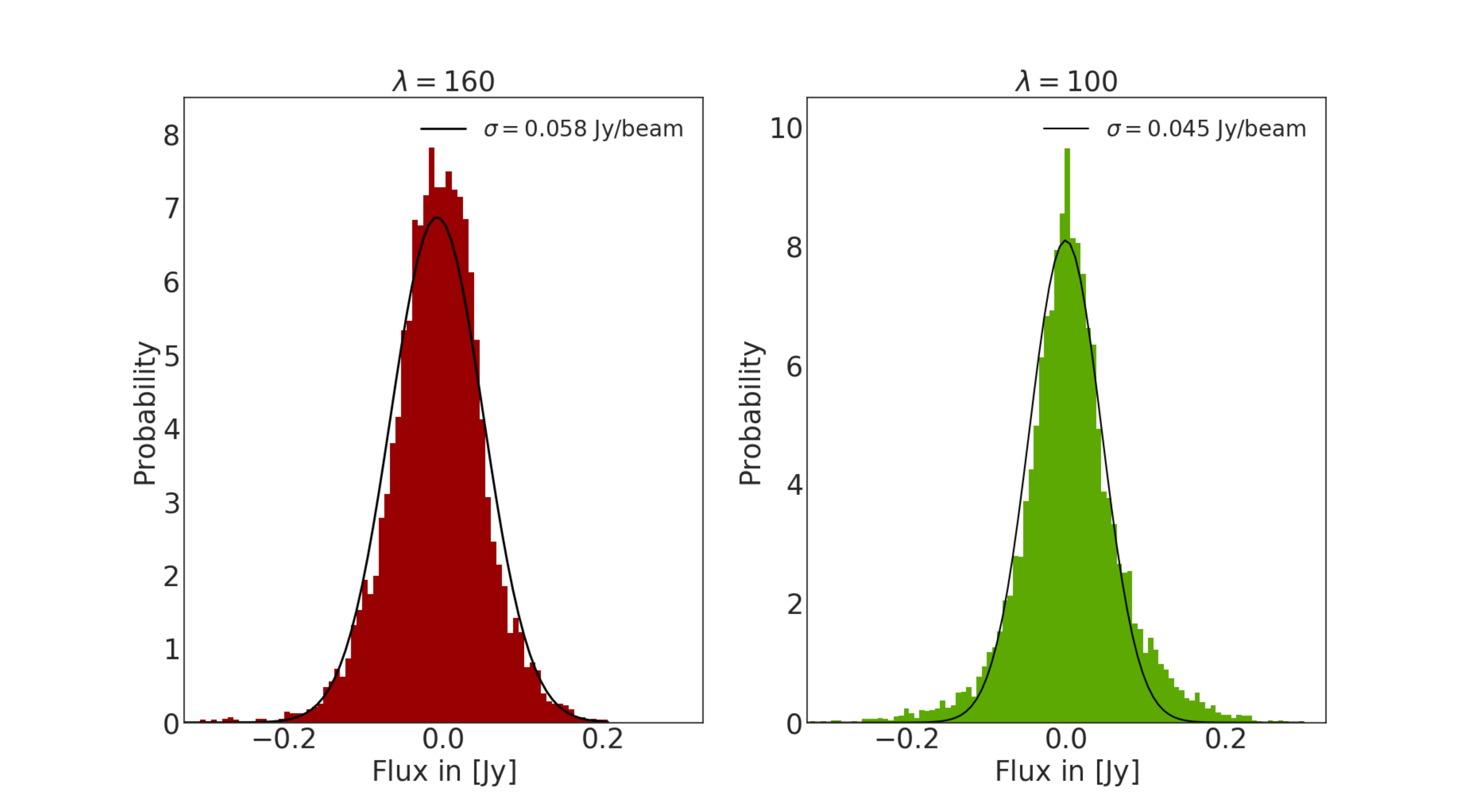}
	\captionof{figure}[Histogram of all measured amplitudes of the reference pixels]{Histogram of all measured amplitudes of the reference pixels during the March 17 observation; the left histogram is for the red band and the right for the green band. The standard deviation is $\sigma = 0.06\mathrm{Jy/beam}$ for the red band and $\sigma = 0.05\mathrm{Jy/beam}$ for the green band.}
	\label{histo}
\end{Figure}
\begin{Figure}
\centering
	\includegraphics[width=\textwidth]{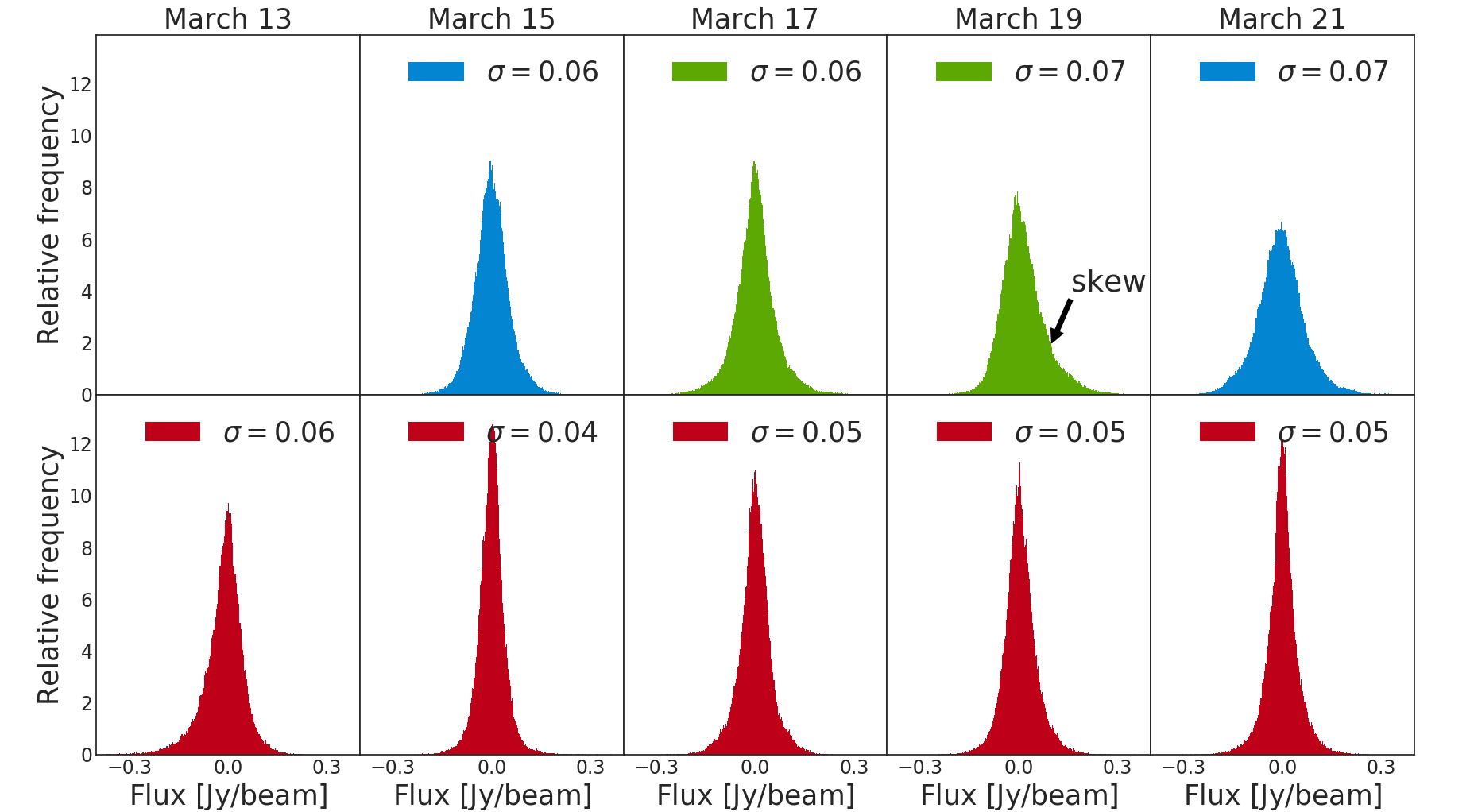}
	\captionof{figure}[Histogram of all measured amplitudes of the reference pixels for all nigths]{As in Figure \ref{histo}, but for all nights. For March 19 the skew induced by the dedicated drift correction is visible, which has not been corrected in this histogram.}
	\label{allHisto}
\end{Figure}

\section{Manual fine tuning}
\label{ARoLC}
\subsection{Manual fine tuning for March 17}
For the March 17 observation, one notices that the flux at the position of Sgr~A* varies more than at the reference points. Inspection shows a discernible point source. Consequently, we only use the first five and the last ten maps to compute the median map and the linear fit. This is a robust method as the linear slope is predominantly constrained by the boundary points and there are still enough ($15$) maps to compute a well-defined median. 

The validity of this can be checked by inspecting the reference light curves: the signal drifts are efficiently removed for all reference light curves. 

We point out that the variability is significant even without this additional step.

\subsection{Manual fine tuning for March 19}
For the March 19 observation, a flux increase occurs during the middle and end times of the observation. This makes a robust correction of the linear drift more difficult. The increase in flux in the middle of the observation is only very weak. It is not clear if including it is reasonable or not. Thus we have no obvious criterion which maps to include for the linear drift correction.

To account for this systematic uncertainty, we test different combinations of maps, which we deem reasonable.  Depending on the linear drift correction we obtain different values of the flux excursions. We estimate the systematic uncertainty as the minimal and maximal value produced with these corrections. For the red band light curve this adds a systematic uncertainty of $\pm0.02~\mathrm{Jy}$ for the peak flux. For the green band the systematic uncertainty is $+0.05~-0.01~\mathrm{Jy}$. The light curves shown in Figure \ref{March19} are for the choice which we consider the most reasonable: The first 14 maps as well as maps $20$ to $30$ determine the linear drift correction. In addition, we neglect the first map of this observation, as a glitch in the reduction rendered it unusable.

As the flux excursion happens during the end of the observation, the linear drift is intrinsically less constrained (because we extrapolate drift for the last maps of this observation based on the previous maps). This manifests itself as an on average increase of the reference light curves at the end of the night. To correct this we subtract the mean of the reference light curves in each map. This is only necessary for this night, as the drift for the other observations is well constrained.

\subsection{Manual fine tuning for other observations}
The light curves of the March 13 and 15 observation show weak excursions (Figure \ref{AllNights}). However, even after the manual fine tuning of the linear drift correction, none of the excursions are significant. The March 21 observation shows no excursion. 

\section{Other observations}
\label{otherNights}
All available light curves are shown in Figure \ref{AllNights}. 
\paragraph{March 13:}
The blue light curve of the first observation, March 13, experiences a 'U-like' drop. We were not able to identify the source of this signal drift nor were we able to correct it. We therefore neglected the blue March 13 observation for all analysis. The parallel red band observation is seemingly unimpaired, however caution is clearly advised. 

\paragraph{March 15:}
There is no significant flux excursion in the blue light curve. 

The flux excursion seen in the red light curve is not significant and we cannot find a discernible point source at the position of Sgr~A*; even after the manual fine tuning. Thus, we cannot claim a detection here and consequently do not use this observation to derive estimates for the SED.

\paragraph{March 21:}
No flux excursions are identifiable in neither of the light curves of this observation. The parallel NIR light curves show weak NIR flares with an intensity comparable to those of the March 17 NIR flares. 

\end{multicols}
\begin{landscape}
\begin{Figure}
\centering	\includegraphics[width=1\textwidth,trim={1.7cm 0 1.75cm 0},clip]{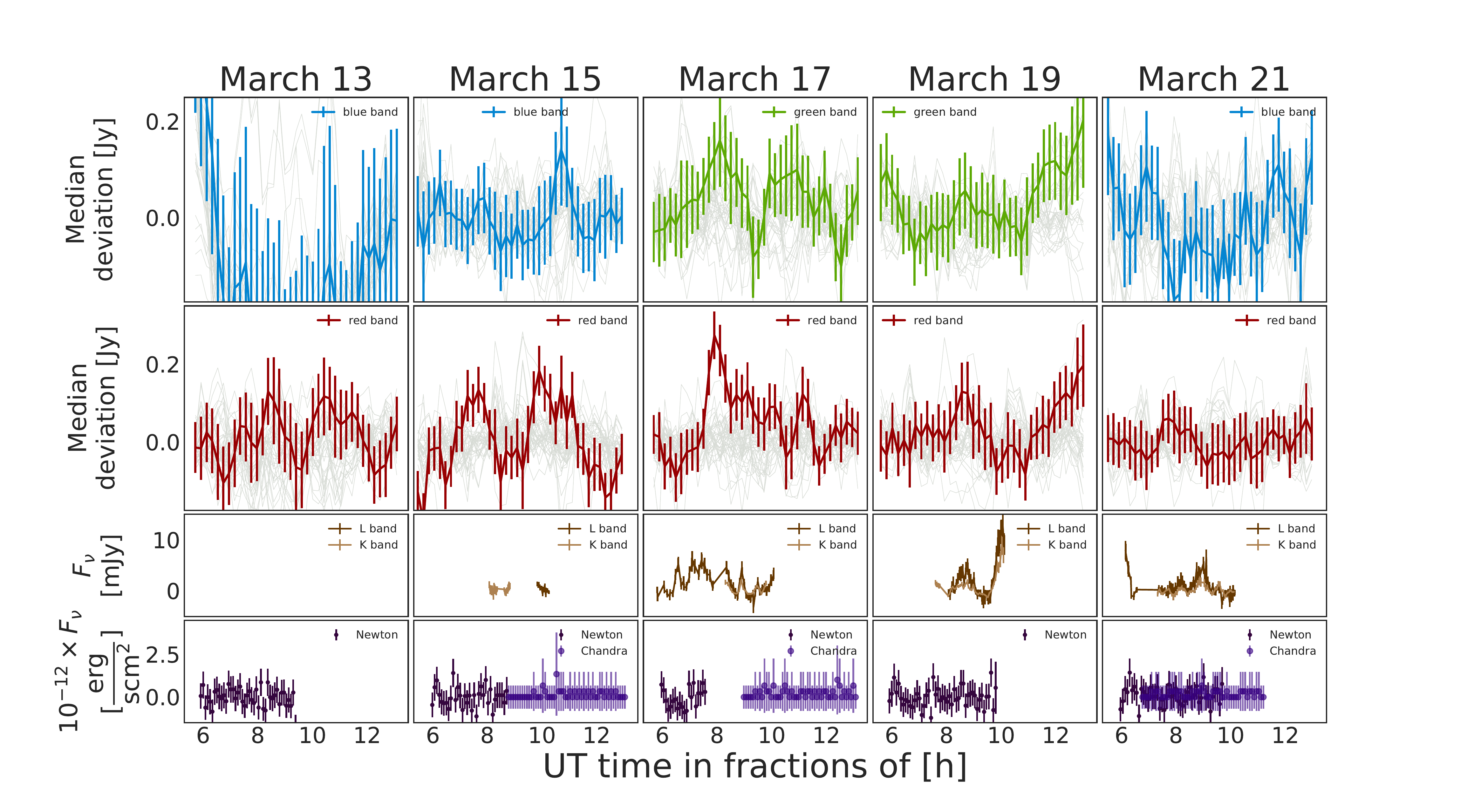}
	\captionof{figure}[All available light curves from our observation campaign]{All availble light curves. The top two panels show the FIR light curves obtained with Herschel/PACS. The top row are the light curves of the blue and green band (color coded). The panel below is the light curve of the red band. The grey light curves are the light curves of the reference positions (see Section \ref{reference_error}). The lower two panels show the NIR (L' and K band) and X-ray parallel observations from NACO, XMM-Newton and Chandra, where available.}
	\label{AllNights}
\end{Figure}
\end{landscape}
\begin{multicols}{2}
\section{Integrated residual maps}
\label{integratedResidualMaps_discussion}
The integrated residual maps show extended flux patches. These are moderately correlated with the regions of high intensity. 

However, this correlation is not perfect. We are thus not able to correct for these artefacts.
 
We argue that these patches not real, but they occur as we reach the sensitivity limit of our data. All regions which show extended flux patches experience a high variance $\sigma^2$. We illustrate this in Figure \ref{residuals_over_sd}, where we plot the integrated residual map of March 17 and the variance map of this observation. For the computation of the variance map we have excluded the 3 maps with the peak flux of Sgr~A*. In the left of Figure \ref{residuals_over_sd} we have circled regions of extended flux patches. In the variance map, these patches clearly stand out. The region of Sgr~A* on the other hand is not effected by such an extended patch.

In addition, the point source visible in the residual maps, as well as in the integrated residual map, is substantially different from these extended flux patches. This is illustrated in Figure \ref{chi_map_march17}. In this figure we plot the red band integrated residual map (left) and a so called $\eta$ map (right). The value of the pixels in the $\eta$ map is defined as follows:
\begin{equation}
\eta_{x,y} = [ \tilde{\chi}_{x,y}^2/ A_{x,y} ] ^{-1}\,\, , 
\end{equation}
where $\tilde{\chi}_{x,y}^2$ is the $\chi^2$ of a PSF fitted to the pixel ${(x,y)}$ and $A_{x,y}$ is the amplitude of the PSF fitted to this value. Therefore, each pixel in the $\eta$ map represents how well a point source with significant flux fits the data. A good fit is characterized by a high value of $\eta$. This is a similar concept to the one used in the StarFinder algorithm \citep{2000SPIE.4007..879D}. 

Inspecting the $\eta$ map reveals that, for the March 17 observation, the only region where we can fit a PSF with a low $\tilde{\chi}^2$ and significant flux is the position of Sgr~A*. 

We repeat this for March 19 and both observations together in Figure \ref{chi_map_allnights}. For March 19 the situation is more ambiguous than for March 17. This reflects the lower significance of the signal.

\end{multicols}

\begin{landscape}
\begin{Figure}
\centering
\includegraphics[width=\textwidth,keepaspectratio,trim={0.cm 0.cm 0cm 0.cm},clip]{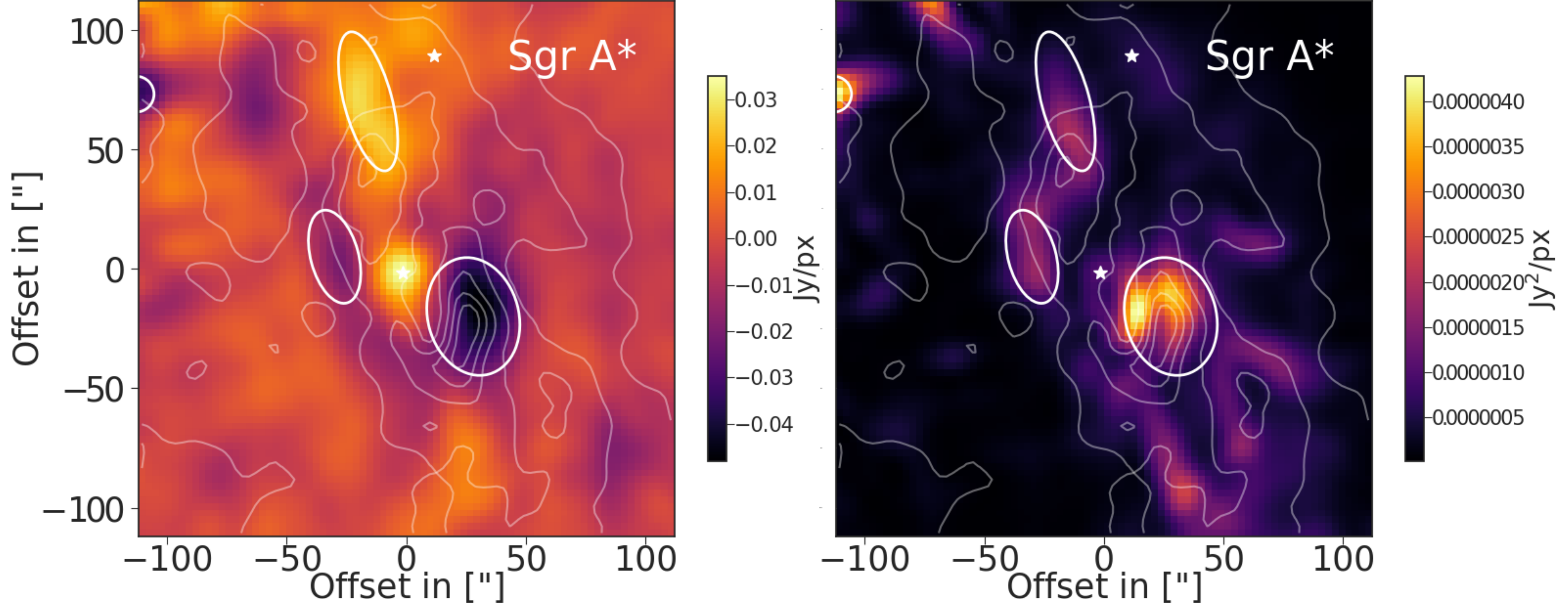}
	\captionof{figure}[Residual maps divided by standard deviation]{Right: red band integrated residual map of March 17. Left: variance map of the residual maps make up the integrated residual map. The extended flux patches in the integrated residual maps have been circled in both plots.}
	\label{residuals_over_sd}
\end{Figure}
\end{landscape}

\begin{landscape}
\begin{Figure}
\centering
	\includegraphics[width=\textwidth,keepaspectratio,trim={0cm 0cm 0cm 0cm},clip]{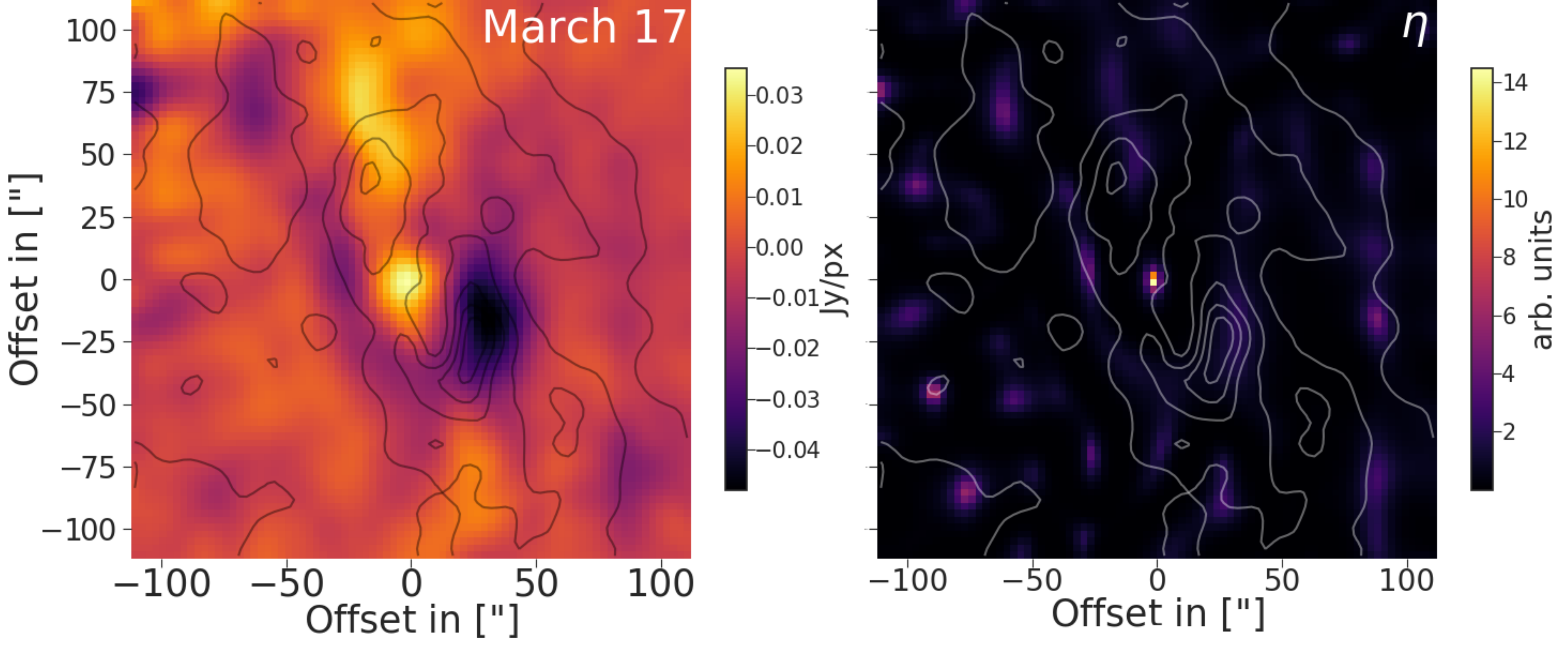}
	\captionof{figure}[Significance of point sources in the integrated residual map]{Significance of point sources in the red band integrated residual map:
	The upper left image shows the integrated residual map of March 17. The map below depicts the inverse of the $\tilde{\chi}^2$ value of a PSF fitted to each pixel of the integrated residual map. The large map to the right shows the same, with the difference that the inverse of the $\tilde{\chi}^2$ value is weighted with the amplitude $A$ of the respective PSF.}
	\label{chi_map_march17}
\end{Figure}

\begin{Figure}
\centering
	\includegraphics[width=\textwidth,keepaspectratio,trim={0cm 0.0cm 0cm 0.cm},clip]{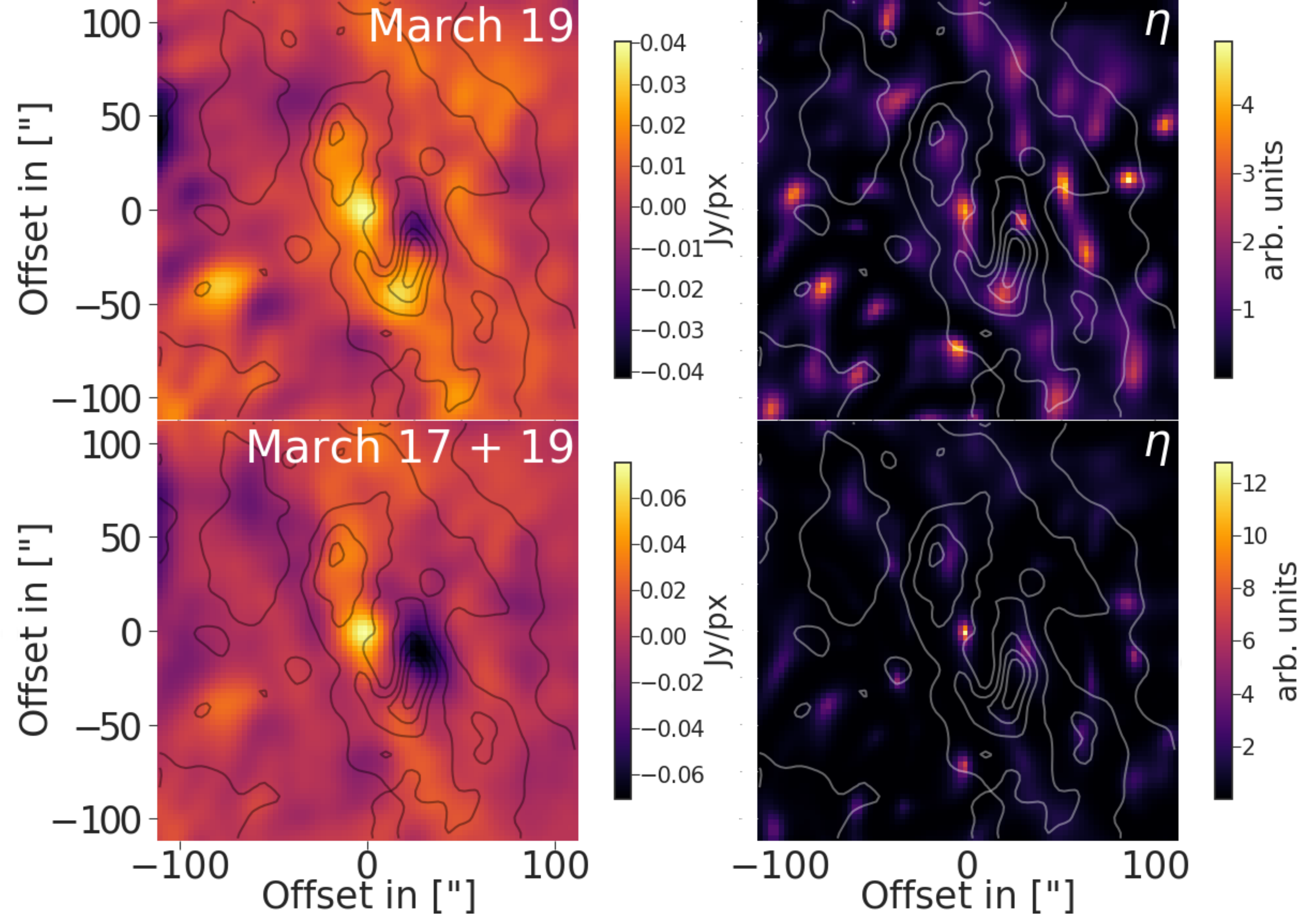}
	\captionof{figure}[Significance of point sources in integrated residual maps]{Significance of a point source in the red band integrated residual maps:
	Similar to Figure \ref{chi_map_march17}, but for the March 19 observation as well as the integrated residual map for both observations, March 17 and 19.}
	\label{chi_map_allnights}
\end{Figure}
\end{landscape}

\end{document}